\documentclass[twocolumn,journal]{IEEEtran}
\usepackage[T1]{fontenc}
\usepackage[latin9]{inputenc}
\usepackage{geometry}
\usepackage{array}
\usepackage{float}
\usepackage{amsmath}
\usepackage{amsthm}
\usepackage{amssymb}
\usepackage{stackrel}
\usepackage{graphicx}
\usepackage[unicode=true,
 bookmarks=true,bookmarksnumbered=true,bookmarksopen=true,bookmarksopenlevel=1,
 breaklinks=false,pdfborder={0 0 0},pdfborderstyle={},backref=false,colorlinks=false]
 {hyperref}
\hypersetup{pdftitle={Your Title},
 pdfauthor={Your Name},
 pdfpagelayout=OneColumn, pdfnewwindow=true, pdfstartview=XYZ, plainpages=false}

\makeatletter

\providecommand{\tabularnewline}{\\}
\floatstyle{ruled}
\newfloat{algorithm}{tbp}{loa}
\providecommand{\algorithmname}{Algorithm}
\floatname{algorithm}{\protect\algorithmname}

\theoremstyle{plain}
\newtheorem{thm}{\protect\theoremname}
\theoremstyle{remark}
\newtheorem{rem}[thm]{\protect\remarkname}
\theoremstyle{plain}
\newtheorem{lem}[thm]{\protect\lemmaname}
\newenvironment{lyxlist}[1]
	{\begin{list}{}
		{\settowidth{\labelwidth}{#1}
		 \setlength{\leftmargin}{\labelwidth}
		 \addtolength{\leftmargin}{\labelsep}
		 }}
	{\end{list}}

\usepackage[caption=false,font=footnotesize]{subfig}
\usepackage{cleveref}
\theoremstyle{remark}

\crefname{lemma}{Lemma}{Lemmas}
\usepackage{cite}
\usepackage[compact]{titlesec}
\usepackage{flushend}

\@ifundefined{showcaptionsetup}{}{%
 \PassOptionsToPackage{caption=false}{subfig}}
\usepackage{subfig}
\makeatother

\providecommand{\lemmaname}{Lemma}
\providecommand{\remarkname}{Remark}
\providecommand{\theoremname}{Theorem}

\begin{document}
\title{Optimal Sensor Placement for Hybrid Source Localization Using Fused
TOA-RSS-AOA Measurements}
\author{Kuntal Panwar, Ghania Fatima\thanks{Kuntal Panwar and Ghania Fatima have contributed equally to the work.},
and Prabhu Babu\thanks{The authors are with CARE, IIT Delhi, New Delhi, 110016, India. (Email:
\{Kuntal.Panwar, Ghania.Fatima, prabhubabu\}@care.iitd.ac.in).}}
\maketitle
\begin{abstract}
Source localization techniques incorporating hybrid measurements improve
the reliability and accuracy of the location estimate. Given a set
of hybrid sensors that can collect combined time of arrival (TOA),
received signal strength (RSS) and angle of arrival (AOA) measurements,
the localization accuracy can be enhanced further by optimally designing
the placements of the hybrid sensors. In this paper, we present an
optimal sensor placement methodology, which is based on the principle
of majorization-minimization (MM), for hybrid localization technique.
We first derive the Cramer-Rao lower bound (CRLB) of the hybrid measurement
model, and formulate the design problem using the A-optimal criterion.
Next, we introduce an auxiliary variable to reformulate the design
problem into an equivalent saddle-point problem, and then construct
simple surrogate functions (having closed form solutions) over both
primal and dual variables. The application of MM in this paper is
distinct from the conventional MM (that is usually developed only
over the primal variable), and we believe that the MM framework developed
in this paper can be employed to solve many optimization problems.
The main advantage of our method over most of the existing state-of-the-art
algorithms (which are mostly analytical in nature) is its ability
to work for both uncorrelated and correlated noise in the measurements.
We also discuss the extension of the proposed algorithm for the optimal
placement designs based on D and E optimal criteria. Finally, the
performance of the proposed method is studied under different noise
conditions and different design parameters.
\end{abstract}

\begin{IEEEkeywords}
Optimal sensor placement, majorization-minimization, saddle-point
formulation, maximum-likelihood estimation, source localization, time
of arrival, angle of arrival, received signal strength, wireless sensor
networks.
\end{IEEEkeywords}

\section{INTRODUCTION AND RELEVANT LITERATURE}

Accurate source localization has been one of the most well studied
techniques due to its wide range of applications in sonars, radars
and wireless sensor networks. A variety of source localization approaches
are available in the literature, which can be classified under the
following categories: time of arrival (TOA) \cite{key-1}, time difference
of arrival (TDOA) \cite{key-2}, received signal strength (RSS) \cite{key-3},
frequency difference of arrival (FDOA) \cite{key-4}, and angle of
arrival (AOA) \cite{key-5}. The aforementioned techniques differ
in the way they use sensor measurements to localize the source, and
they also differ in their applicability depending on the range of
the source to be localized. For instance, techniques like TOA and
TDOA are generally preferred for long range source localization and
RSS is usually preferred for short range localization. In the recent
years, various hybrid source localization techniques, which are based
on the combination of the aforementioned techniques, are proposed
and are observed to be range-independent. Some of the hybrid source
localization techniques available in the literature are as follows:
TOA-RSS \cite{key-6,key-7,key-8,key-9,key-10}, RSS-AOA \cite{key-11,key-12,key-13,key-14,key-15},
TOA-AOA \cite{key-16},TDOA-AOA \cite{key-17,key-18,key-19} and AOA-TDOA-FDOA
\cite{key-20,key-21}. As the modern-day sensors collecting the measurements
are generally capable to collect more than one type of measurements
(they are capable of collecting both range and angle measurements),
one can carefully design their placements which will lead to improved
localization accuracy \cite{key-22,key-23,key-24}. 

The relation between the sensor locations and the Cramer-Rao lower
bound (CRLB) matrix of the localization model has been well studied
\cite{key-25} and the optimal positions of sensors can indeed be
designed by optimizing a related metric of CRLB matrix. Based on the
choice of the design metric, the optimal sensor placement techniques
can be categorized as \cite{key-26}: A-optimal design (minimizing
the trace of CRLB matrix), D-optimal design (minimizing the determinant
of CRLB matrix), and E-optimal design (minimizing the maximum eigenvalue
of CRLB matrix). The most commonly used criterion is the A-optimal
criterion since it denotes the average variance of the estimates,
and therefore, minimizing the A-optimal criterion is equivalent to
minimization of the mean square error (MSE).

A wide variety of literature on optimal sensor placement approaches
for individual source localization techniques are available. For instance,
in \cite{key-27}, the optimal sensor placement for target localization
by bearing measurements was obtained via maximization of determinant
of Fisher information matrix (FIM). The authors in \cite{key-28}
have derived the CRLB for sensor placement for TDOA measurements and
found the optimal geometry to be centered platonic solids (cube, tetrahedron,
etc) with target at its center and sensors at each vertices. The D-optimal
criterion was used for obtaining placement strategies for localization
using range measurements in \cite{key-29,key-30}, received signal
strength difference measurements (RSSD) in \cite{key-31} and TOA
measurements in \cite{key-32,key-33}. The estimation accuracy of
elliptical TOA localization was improved using minimizing the trace
of the CRLB in \cite{key-34}. In \cite{key-35} and \cite{key-36},
the optimal geometry is derived using a generalized inequality for
the A-optimal criterion for AOA measurements. This approach was further
extended to RSS in \cite{key-37} and TOA in \cite{key-38}. The authors
in \cite{key-39} derived the optimal sensing directions for target
localization for different types of measurements while considering
the uncertainty in the prior knowledge of the target position. A unified
framework for designing optimal sensor orientations based on A, D
and E optimal criteria was developed for different types of measurement
models in \cite{key-40}.

\begin{table*}[t]
\centering \caption{{\footnotesize{}COMPARISON BETWEEN RELATED PREVIOUS STUDIES AND PROPOSED
ALGORITHM}}
\label{theory} %
\begin{tabular}{ccccc}
\hline 
\textbf{\small{}Paper} & \textbf{\small{}Hybrid measurements} & \textbf{\small{}Correlated noise} & \textbf{\small{}Non-uniform noise variances} & \textbf{\small{}Optimality criteria}\tabularnewline
\hline 
{\small{}\cite{key-28}} & {\small{}No} & {\small{}No} & {\small{}No} & {\small{}A}\tabularnewline
{\small{}\cite{key-22,key-24,key-27,key-29,key-30}} & {\small{}No} & {\small{}No} & {\small{}No} & {\small{}D}\tabularnewline
{\small{}\cite{key-34,key-35,key-36,key-37,key-38}} & {\small{}No} & {\small{}No} & {\small{}Yes} & {\small{}A}\tabularnewline
{\small{}\cite{key-31,key-32,key-33}} & {\small{}No} & {\small{}No} & {\small{}Yes} & {\small{}D}\tabularnewline
{\small{}\cite{key-40}} & {\small{}No} & {\small{}Yes} & {\small{}Yes} & {\small{}A, D and E}\tabularnewline
{\small{}\cite{key-39,key-41}} & {\small{}Yes} & {\small{}No} & {\small{}No} & {\small{}A, D}\tabularnewline
{\small{}\cite{key-42}} & {\small{}Yes} & {\small{}No} & {\small{}No} & {\small{}D}\tabularnewline
{\small{}\cite{key-23,key-43}} & {\small{}Yes} & {\small{}No} & {\small{}No} & {\small{}A}\tabularnewline
{\small{}Proposed Algorithm} & {\small{}Yes} & {\small{}Yes} & {\small{}Yes} & {\small{}A, D and E}\tabularnewline
\hline 
\end{tabular}
\end{table*}

In the case of hybrid localization methods, very few methods are available
to design optimal sensor placements; this maybe due to complicated
dependence of the hybrid CRLB matrix in terms of the sensor locations.
In \cite{key-41}, optimal velocity configurations were derived for
a sensor-target geometry to localize a stationary emitter using TDOA-FDOA
measurements. Using D-optimal design, both the sensor placements and
velocity configurations were derived in \cite{key-42} for hybrid
TDOA-FDOA sensors. The authors of \cite{key-43} have designed optimal
sensor placement strategies by considering A optimal design for the
hybrid TOA-RSS-AOA model; however, their approach is restrictive as
they assume the noise in the measurements to be uncorrelated with
uniform noise variance, and moreover the design metric is limited
only to A-optimal criterion. A brief summary of the various works
from the literature on optimal sensor placement techniques is given
in Table \ref{theory}. As noted above, most of the aforementioned
methods for the hybrid model invariably assume the noise in the measurements
to be uncorrelated and uniform, which is quite restrictive as there
are many applications, like underwater surveillance \cite{key-44},
where the noise in the sensor measurements are correlated, and in
almost all real-life scenarios the noise variance in the measurements
will not be equal. So, there is a need to devise a method that can
effectively design optimal sensor geometries in the case of hybrid
localization models for both correlated and uncorrelated noise (uniform
and non-uniform variances) and include all the optimal designs (A,
D and E).

In this paper, for the hybrid TOA-RSS-AOA measurement model, an optimal
sensor placement algorithm which can optimize all the three design
criteria and handle correlated noise measurements has been proposed.
The associated non-convex design problem is first reformulated using
either the Fenchel or the Lagrangian formulation into a saddle-point
problem. The equivalent saddle-point problem is then solved using
the principle of majorization-minimization (MM), which is applied
over both primal and dual variables. The proposed algorithm can work
efficiently for all types of noise models.

We would like to note that the ADMM based approach proposed in \cite{key-40}
is not extendable for hybrid source localization model, as extending
the framework in \cite{key-40} to handle the hybrid CRLB is not straightforward
due to the presence of additional rotation matrix in the FIM of the
AOA model. On the other hand, the proposed algorithm can easily handle
the design objective based on CRLB for the hybrid measurement model.

The novelty and the key contributions of our work are listed as follows: 
\begin{itemize}
\item We derive the CRLB of the hybrid TOA-RSS-AOA model. The CRLB is derived
for general non-diagonal noise covariance matrices (assuming the noise
in the measurements to be correlated). Although the derivation of
CRLB for the hybrid model is standard and can be found in the literature,
unlike the commonly used parametrization of expressing the elements
of CRLB in terms of highly non-linear trignometric functions, we parameterize
the elements of the CRLB matrix in terms of some unit norm vectors
which will ease the complicacy of the design criterion and help us
to devise a numerical algorithm. The aforementioned reparametrization
can swiftly handle the variations in the expressions of the CRLB matrices
associated with different measurement models. 
\item We then formulate the design objective by considering A, D and E optimal
criteria based on the hybrid CRLB. For each of the design objective,
we derive a monotonic algorithm based on the MM principle. The MM
technique developed in this paper is novel and unlike the conventional
MM methods, which work only on the primal optimization variable, it
exploits both primal and dual variable updates in its steps. More
precisely, we reformulate each of the optimal design problem (A, D
and E) into a saddle-point problem and design novel upperbounds over
primal and dual variables. The general framework of our primal-dual
MM algorithm can inspire researchers to solve many different optimization
problems where devising the conventional MM is not easy. 
\item We discuss the computational complexities of the proposed algorithm
for the three design problems, and discuss the proof of convergence.
\item We present extensive numerical simulations under different noise models
(correlated, uncorrelated-uniform and non-uniform) to prove the effectiveness
of proposed method in designing optimal sensor locations for the hybrid
model. We also present studies based on MSE for the optimal orientations
obtained via our approach.
\end{itemize}
The rest of the paper is organized as follows: Section II discusses
the data model for the hybrid measurements and formulates the design
problem for optimal sensor placement. Section III gives a brief overview
of the MM principle, introduces the proposed MM based algorithm, discusses
the computation complexity and convergence of the proposed algorithm,
and presents its extensions to deal with D and E optimality criteria.
Section IV gives the details of numerical simulations and the results.
Finally, Section V concludes the paper and mentions possible future
directions.

\textit{Notations:} Vectors, matrices and scalars are represented
by bold lowercase letters (e.g., $\mathbf{a}$), bold uppercase letters
(e.g., $\mathbf{A}$) and italic letters (e.g., $a$, $A$), respectively.
The $i^{\text{th}}$ element of vector $\mathbf{a}$ is denoted as
$a_{i}$. The $i^{\text{th}}$ column and the $(i,j)^{\text{th}}$
element of the matrix $\mathbf{A}$ are denoted as $\mathbf{a}_{i}$
and $A_{i,j}$, respectively. For a vector $\mathbf{a}$, its Euclidean
norm is denoted as $\left\Vert \mathbf{a}\right\Vert $. The transpose,
inverse, square and square root of a matrix are defined as $\mathbf{A}^{T}$,
$\mathbf{A}^{-1}$, $\mathbf{A}^{2}$, and $\sqrt{\mathbf{A}}$ respectively.
The notations $\textrm{Tr}(\mathbf{A}),\;\lambda_{\mathrm{max}}(\mathbf{A}),\;\lambda_{\mathrm{min}}(\mathbf{A}),\;\mathrm{and}\;\|\mathbf{A}\|_{F}$
represent trace, maximum eigenvalue, minimum eigenvalue and Frobenius
norm of a matrix, respectively. $\text{ln}(\cdot)$ represents natural
logarithm and $\mathbf{I}_{m}$ represents identity matrix of size
$m\times m$. The subscript $\mathbf{a}_{t}$ denotes the estimate
of the vector \textbf{$\mathbf{a}$} at the $t^{\mathrm{th}}$ iteration.
A diagonal matrix is represented as $\mathrm{diag}(\mathbf{a})$,
where $\mathbf{a}$ is the vector containing the diagonal elements,
and a block diagonal matrix is represented as $\mathrm{blkdiag}\left(\mathbf{A},\mathbf{B}\right)$,
where $\mathbf{A}$ and $\mathbf{B}$ are the smaller sub-matrices.
$\mathbf{S}_{+}^{n}$ denotes the set of symmetric positive semidefinite
matrices of size $n\times n$ and $\mathbb{N}$ denotes the set of
natural numbers. The notations $\mathcal{N}\left(\mu,\sigma^{2}\right)$
denotes Gaussian distribution with mean $\mu$ and variance $\sigma^{2}$,
$\mathcal{U}\left(a,b\right)$ denotes uniform distribution taking
values between $a$ and $b$ and $p\left(\mathbf{x}\right)$ denotes
the probability density function of the random variable $\mathbf{x}$.

\section{PROBLEM FORMULATION}

In this section, we formulate the optimization problem for optimal
sensor placement to accurately localize a static target using hybrid
measurements from $m$ stationary sensors in an $n$-dimensional space.
For convenience, we present the derivation for the two-dimensional
(2D) case and mention the necessary changes to be adopted for the
three-dimensional (3D) case as a separate remark later. From hereon,
we will use the terms source and target interchangeably.

\begin{figure}[tbh]
\begin{centering}
\includegraphics[scale=0.15]{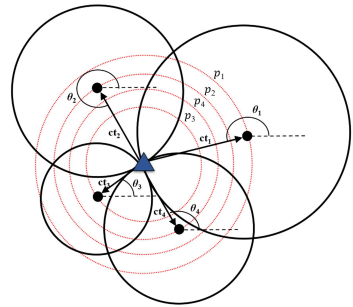}
\par\end{centering}
\caption{{\small{}\label{fig:1}Measurement geometry of the hybrid localization
model. The blue triangle represents the target and the black circles
are the sensors. $p_{1}$, $p_{2}$, $p_{3}$ and $p_{4}$ represent
the RSS measurements, $\theta_{1}$, $\theta_{2}$, $\theta_{3}$
and $\theta_{4}$ represent the AOA measurements, and $ct_{1}$, $ct_{2}$,
$ct_{3}$ and $ct_{4}$ are the measurements corresponding to TOA.}}
\end{figure}

Fig. \ref{fig:1} shows the hybrid localization geometry in which
four sensors collect TOA, RSS and AOA measurements of a source. Let
the position coordinates of the source/target and the $i^{\mathrm{th}}$
sensor be denoted by $\mathbf{r}=[r_{x},r_{y}]^{T}$ and $\mathbf{s}_{i}=[s_{xi},s_{yi}]^{T}$,
respectively. Assuming each sensor obtain its corresponding TOA, RSS
and AOA measurements, the TOA measurement at the $i^{\text{th}}$
sensor is given as follows: 
\begin{equation}
t_{i}=\frac{\|\mathbf{r}-\mathbf{s}_{i}\|}{c}+\mu_{i},\;\forall i=1,\dots,m,\label{time}
\end{equation}
where $c$ is the speed of the signal and $d_{i}\triangleq\|\mathbf{r}-\mathbf{s}_{i}\|$
is the distance between the $i^{\text{th}}$ sensor and the target.
The measurements are assumed to be corrupted by zero-mean Gaussian
noise ($\mu_{i}$). Computing the corresponding range measurements
from \eqref{time}, we have 
\begin{equation}
z_{i}=d_{i}+c\mu_{i},\;\forall i=1,\ldots,m.
\end{equation}
Integrating the TOA measurements from $m$ sensors, we get the following
measurement model: 
\begin{equation}
\mathbf{z}=\mathbf{\mathbf{g}}_{\text{TOA}}(\mathbf{r})+\mathbf{\boldsymbol{\xi}}_{\text{TOA}}\label{eq:3}
\end{equation}
where $\mathbf{z}=[z_{1},\dots,z_{m}]^{T}$ are the measurements,
$\mathbf{g}_{\mathrm{TOA}}(\mathbf{r})=\left[d_{1},\ldots,d_{m}\right]^{T}$
and the error vector is $\mathbf{\boldsymbol{\xi}}_{\text{TOA}}=[c\mu_{1},\dots,c\mu_{m}]^{T}\sim\mathcal{N}(\boldsymbol{0},\boldsymbol{\Sigma}_{\text{TOA}})$.\\

The RSS measurement (average power $p_{i}$ (in dB)) received at the
$i^{\text{th}}$ sensor is given as follows \cite{key-45}: 
\begin{equation}
p_{i}=p_{0}-10\alpha\ln\left(d_{i}\right)+\nu_{i},\;\forall i=1,\dots,m,\label{rss_sys}
\end{equation}
where $p_{0}$ is the signal power at the target and $\alpha$ is
the path loss constant which depends on the medium. It is assumed
that the parameters $p_{0}$ and $\alpha$ are known (determined using
calibration \cite{key-45}). The noise ($\nu_{i}$) in the measurements
is assumed to be Gaussian noise. Integrating the RSS measurements
from $m$ sensors we get: 
\begin{equation}
\mathbf{p}=\eta\,\mathbf{g}_{\mathrm{RSS}}(\mathbf{r})+\mathbf{\boldsymbol{\xi}}_{\text{RSS}},\label{eq:5}
\end{equation}
where $\mathbf{p}=[p_{1}-p_{0},\dots,p_{m}-p_{0}]^{T}$, $\eta=\frac{-10\alpha}{\mathrm{ln}(10)}$,
$\mathbf{g}_{\mathrm{RSS}}(\mathbf{r})=[\ln(d_{1}),\dots,\ln(d_{m})]^{T}$
and $\mathbf{\boldsymbol{\xi}}_{\text{RSS}}=[\nu_{1},\dots,\nu_{m}]^{T}\sim\mathcal{N}(\boldsymbol{0},\boldsymbol{\Sigma}_{\text{RSS}})$.

The AOA measurement at the $i^{\text{th}}$ sensor is given by: 
\begin{equation}
\theta_{i}=\tan^{-1}\left(\frac{r_{y}-s_{yi}}{r_{x}-s_{xi}}\right)+\omega_{i},\;\forall i=1,\dots,m,
\end{equation}
where $\tan^{-1}\left(\cdot\right)$ is the 4-quadrant arctangent.
If we assume the measurements are corrupted by Gaussian noise ($\omega_{i}$),
then the AOA measurement model for $m$ sensors is given as: 
\begin{equation}
\mathbf{\boldsymbol{\mathbf{\theta}}}=\mathbf{g}_{\mathrm{AOA}}(\mathbf{r})+\mathbf{\boldsymbol{\xi}}_{\text{AOA}},\label{eq:7}
\end{equation}
where $\boldsymbol{\mathbf{\theta}}=[\theta_{1},\dots,\theta_{m}]^{T}$,
$\mathbf{g}_{\mathrm{AOA}}(\mathbf{r})=[\tan^{-1}\left(\frac{r_{y}-s_{y1}}{r_{x}-s_{x1}}\right),\dots,\tan^{-1}\left(\frac{r_{y}-s_{ym}}{r_{x}-s_{xm}}\right)]^{T}$
and $\boldsymbol{\mathbf{\xi}}_{\text{AOA}}=[\omega_{1},\dots,\omega_{m}]^{T}\sim\mathcal{N}(\boldsymbol{0},\boldsymbol{\Sigma}_{\text{AOA}})$.

With the individual data models in (\ref{eq:3}), (\ref{eq:5}) and
(\ref{eq:7}), the hybrid TOA-RSS-AOA measurement model can be obtained
by stacking the individual models as given below:
\begin{equation}
\mathbf{q}=\begin{bmatrix}\mathbf{z}\\
\mathbf{p}\\
\mathbf{\boldsymbol{\theta}}
\end{bmatrix}=\begin{bmatrix}\mathbf{\mathbf{\mathbf{g}}_{\text{TOA}}(\mathbf{r})}\\
\eta\,\mathbf{\mathbf{g}}_{\text{RSS}}(\mathbf{r})\\
\mathbf{\mathbf{g}}_{\text{AOA}}(\mathbf{r})
\end{bmatrix}+\begin{bmatrix}\mathbf{\mathbf{\boldsymbol{\xi}}_{\text{TOA}}}\\
\mathbf{\boldsymbol{\xi}}_{\text{RSS}}\\
\mathbf{\boldsymbol{\xi}}_{\text{AOA}}
\end{bmatrix}.
\end{equation}
The FIM matrix for the hybrid model is defined as:
\begin{equation}
\mathbf{F}_{\mathrm{Hybrid}}=\mathbb{E}\left(\left(\frac{\partial\ln p(\mathbf{q};\mathbf{r})}{\partial\mathbf{r}}\right)\left(\frac{\partial\ln p(\mathbf{q};\mathbf{r})}{\partial\mathbf{r}}\right)^{T}\right)\in\mathbb{R}^{n\times n},\label{eq:9}
\end{equation}
where $p(\mathbf{q};\mathbf{r})$ is the joint probability density
function of the hybrid model. If we assume the noise in the three
localization models ($\mathbf{\boldsymbol{\xi}}_{\text{TOA}}$, $\mathbf{\boldsymbol{\xi}}_{\text{RSS}}$
and $\mathbf{\boldsymbol{\xi}}_{\text{AOA}}$) are independent\footnote{Every hybrid sensor comprises of different mechanisms to collect the
TOA, RSS and AOA measurements. Therefore, it is natural to assume
the noise in the different measurements to be independent of each
other. However, the noise in individual measurements may be correlated,
which we have already modelled by taking the individual noise covariance
matrices to be non-diagonal.} of each other, the joint probability density function will be proportional
to the product of their individual probability density functions:
\begin{equation}
p\left(\mathbf{q};\mathbf{r}\right)\propto p\left(\mathbf{z};\mathbf{r}\right)\times p\left(\mathbf{p};\mathbf{r}\right)\times p\left(\mathbf{\boldsymbol{\theta}};\mathbf{r}\right).
\end{equation}
The FIM for the hybrid measurements can therefore be written as \cite{key-36}:
\begin{equation}
\mathbf{F}_{\mathrm{Hybrid}}(\mathbf{r})=\mathbf{F}_{\text{TOA}}(\mathbf{r})+\mathbf{F}_{\text{RSS}}(\mathbf{r})+\mathbf{F}_{\text{AOA}}(\mathbf{r}),
\end{equation}
where $\mathbf{F}_{\text{TOA}}$, $\mathbf{F}_{\text{RSS}}$ and $\mathbf{F}_{\text{AOA}}$
are the individual FIM matrices and are given by (\ref{eq:9}) with
the joint probability density function replaced by their individual
probability density functions. 

We first derive the CRLB matrix for TOA measurement. The probability
density function of the observation vector $\mathbf{z}$ for the TOA
model is given as: 
\begin{equation}
\begin{split}p(\mathbf{z};\mathbf{r})= & \frac{\exp\left(-\frac{1}{2}(\mathbf{z}-\mathbf{g}_{\mathrm{TOA}}(\mathbf{r}))^{T}\boldsymbol{\Sigma}_{\text{TOA}}^{-1}(\mathbf{z}-\mathbf{g}_{\mathrm{TOA}}(\mathbf{r}))\right)}{(2\pi)^{\frac{m}{2}}\sqrt{\det(\boldsymbol{\Sigma}_{\text{TOA}})}}.\end{split}
\end{equation}
Let $\hat{\mathbf{r}}$ be an unbiased estimator of the true target
position, then the covariance matrix satisfies the following well-known
inequality \cite{key-46}: 
\begin{equation}
\mathbb{E}\left[(\hat{\mathbf{r}}-\mathbf{r})(\hat{\mathbf{r}}-\mathbf{r})^{T}\right]\succeq\mathbf{C}_{\text{TOA}}(\mathbf{r})=\mathbf{F}_{\text{TOA}}^{-1}(\mathbf{r}),
\end{equation}
where $\mathbf{C}_{\text{TOA}}(\mathbf{r})$ is the CRLB matrix and
is given as: 
\begin{equation}
\textrm{\textbf{C}}_{\text{TOA}}(\mathbf{r})=\textrm{\textbf{F}}_{\text{TOA}}^{-1}(\mathbf{r})=(\mathbf{J}^{T}\boldsymbol{\mathbf{\Sigma}}_{\text{TOA}}^{-1}\mathbf{J})^{-1},\label{eq:fimtoa}
\end{equation}
where 
\begin{equation}
\textbf{J}=\begin{bmatrix}\cos\theta_{1} & \sin\theta_{1}\\
\vdots & \vdots\\
\cos\theta_{m} & \sin\theta_{m}
\end{bmatrix}=\begin{bmatrix}\frac{(\mathbf{r}-\mathbf{s}_{1})^{T}}{\|\mathbf{r}-\mathbf{s}_{1}\|}\\
\vdots\\
\frac{(\mathbf{r}-\mathbf{s}_{m})^{T}}{\|\mathbf{r}-\mathbf{s}_{m}\|}
\end{bmatrix}.\label{eq:15}
\end{equation}
 The matrix \textbf{J} is referred to as the orientation matrix and
each row of this matrix denotes the unit vector pointing in the direction
of the line connecting the $i^{\mathrm{th}}$ sensor and the target. 

From hereon, we will express the CRLB (or the FIM) matrix only in
terms of $\mathbf{J}$ (as $\mathbf{C}_{\mathrm{TOA}}\left(\mathbf{J}\right)$)
as it can be seen from (\ref{eq:15}) that the rows of $\mathbf{J}$
are only dependent on the orientation and not the actual position
of the sensors or the target.
\begin{rem}
The parameterization of orientation matrix \textbf{J} in (\ref{eq:15})
is stated in both the cartesian and polar forms. Most of the literature
utilizes the polar form, which is specified in terms of `azimuth'
angle, thereby resulting in complex trigonometric expressions for
the elements of the CRLB matrix. In order to simplify the design problem
associated with the CRLB matrix, we will prefer the elements of the
$\mathbf{J}$ to be in the cartesian coordinates. We can easily derive
the respective azimuth angles from the cartesian representation, if
required.
\end{rem}
Similar to the derivation of $\textrm{\textbf{C}}_{\text{TOA}}$,
the CRLB for RSS and AOA can be derived, and their FIM are given by
\cite{key-40}: 
\begin{align}
\mathbf{F}_{\text{RSS}}(\mathbf{J})=\eta^{2}\textbf{J}^{T}\textbf{D}\mathbf{\Sigma}_{\text{RSS}}^{-1}\textbf{D}\mathbf{J},\label{eq:fimrss}\\
\mathbf{F}_{\text{AOA}}(\mathbf{J})=\mathbf{U}^{T}\textbf{J}^{T}\textbf{D}\mathbf{\Sigma}_{\text{AOA}}^{-1}\textbf{D}\mathbf{J}\mathbf{U},\label{eq:fimaoa}
\end{align}
where $\textbf{D}$ is the distance matrix and $\textbf{U}$ is the
unitary matrix which are given by: 
\begin{equation}
\begin{array}{ll}
\textbf{D}=\mathrm{diag}\left(1/d_{1},\ldots1/d_{m}\right)\mathrm{;\;}\textbf{U}=\begin{bmatrix}0 & 1\\
-1 & 0
\end{bmatrix}.\end{array}
\end{equation}

\begin{rem}
\label{rem:3}The CRLB for RSS and AOA depends on the distance matrix
\textbf{D} and the CRLB of the AOA model also involves an extra unitary
matrix \textbf{U} (however, its elements are known). Before venturing
into deriving the CRLB for the hybrid measurement model, it is important
to note that any design objective which is based on the CRLB (for
RSS and AOA) will depend on the target-sensor distances ($d_{i}$)
(which is not known). So, a usual approach is to perform the optimal
placement design based on an initial coarse target estimate. This
initial target position (which maybe very rough) is assumed to be
known by some alternate means and using which the orientation of the
sensors are obtained. Although, this design of optimal placements
by using a rough target location looks heuristic, it works very well
in practice even if there is a slight mismatch between assumed target
location in design and the true underlying unknown target location.
In the numerical section, we will present MSE analysis for this mismatch
issue. 
\end{rem}
With the individual FIMs mentioned in equations (\ref{eq:fimtoa}),
(\ref{eq:fimrss}) and (\ref{eq:fimaoa}), the FIM of the hybrid measurements
($\boldsymbol{\textrm{F}}_{\mathrm{Hybrid}}$) is given by: 
\begin{equation}
\begin{aligned}\mathbf{F}_{\mathrm{Hybrid}}(\mathbf{J})= & \mathbf{J}^{T}\boldsymbol{\mathbf{\Sigma}}_{\text{TOA}}^{-1}\mathbf{J}+\eta^{2}\textbf{J}^{T}\textbf{D}\mathbf{\Sigma}_{\text{RSS}}^{-1}\textbf{D}\mathbf{J}\\
 & +\mathbf{U}^{T}\textbf{J}^{T}\textbf{D}\mathbf{\Sigma}_{\text{AOA}}^{-1}\textbf{D}\mathbf{J}\mathbf{U}.
\end{aligned}
\end{equation}
and the corresponding hybrid CRLB matrix is given as: 
\begin{equation}
\begin{aligned}\textrm{\textbf{C}}_{\mathrm{Hybrid}}(\mathbf{J})= & \left(\mathbf{J}^{T}\boldsymbol{\mathbf{\Sigma}}_{\text{TOA}}^{-1}\mathbf{J}+\eta^{2}\textbf{J}^{T}\textbf{D}\mathbf{\Sigma}_{\text{RSS}}^{-1}\textbf{D}\mathbf{J}\right.\\
 & \left.+\mathbf{U}^{T}\textbf{J}^{T}\textbf{D}\mathbf{\Sigma}_{\text{AOA}}^{-1}\textbf{D}\mathbf{J}\mathbf{U}\right)^{-1}
\end{aligned}
\end{equation}
We would like to note that the hybrid CRLB matrix is the function
of the orientation matrix, and we can optimize a scalar metric of
CRLB to come up with optimal orientation matrix. We first consider
the A-optimal criterion, where the objective is to minimize the trace
of the CRLB matrix which results in the following optimization problem:
\begin{equation}
\begin{aligned}\min\limits _{\textbf{J}} & \;\textrm{Tr}\Biggl(\biggl(\textbf{J}^{T}\mathbf{\Sigma}_{\text{TOA}}^{-1}\textbf{J}+\textbf{J}^{T}\eta^{2}\textbf{D}\mathbf{\Sigma}_{\text{RSS}}^{-1}\textbf{D}\textbf{J}\\
 & \left.\left.\,\,\,\,\,\,\,\,\,\,\,\,\,\,\,\,\,\,\,\,\,\,+\textbf{U}^{T}\textbf{J}^{T}\textbf{D}\mathbf{\Sigma}_{\text{AOA}}^{-1}\textbf{D}\textbf{J}\textbf{U}\right)^{-1}\right)\\
\text{s.t. } & \;\mathbf{j}_{i}^{T}\mathbf{j}_{i}=1,i=1,\cdots,m,
\end{aligned}
\label{eq:20-1}
\end{equation}
where $\mathbf{j}_{i}$ is the $i^{\mathrm{th}}$ column of $\mathbf{J}^{T}$.
Problem (\ref{eq:20-1}) can be rewritten as:
\begin{equation}
\begin{aligned}\min\limits _{\textbf{J}\in\mathcal{D}} & \;\textrm{Tr}\Biggl(\biggl(\textbf{J}^{T}\left(\mathbf{\Sigma}_{\text{TOA}}^{-1}+\eta^{2}\textbf{D}\mathbf{\Sigma}_{\text{RSS}}^{-1}\textbf{D}\right)\textbf{J}\\
 & \left.\left.\,\,\,\,\,\,\,\,\,\,\,\,\,\,\,\,\,\,\,\,\,\,+\textbf{U}^{T}\textbf{J}^{T}\textbf{D}\mathbf{\Sigma}_{\text{AOA}}^{-1}\textbf{D}\textbf{J}\textbf{U}\right)^{-1}\right),
\end{aligned}
\label{eq:21-1}
\end{equation}
where the set $\mathcal{D}$ denotes the constraint $\textbf{j}_{i}^{T}\textbf{j}_{i}=1~\forall~i$.
Using $\textbf{H}=\left[\textbf{J}^{T},\textbf{U}^{T}\textbf{J}^{T}\right]^{T}$
and $\textbf{R}=\text{\text{blkdiag}}\left(\mathbf{R}_{1},\mathbf{R}_{2}\right)$,
where $\mathbf{R}_{1}=\mathbf{\Sigma}_{\text{TOA}}^{-1}+\eta^{2}\textbf{D}\mathbf{\Sigma}_{\text{RSS}}^{-1}\textbf{D}$
and $\mathbf{R}_{2}=\textbf{D}\mathbf{\Sigma}_{\text{AOA}}^{-1}\textbf{D}$,
(\ref{eq:21-1}) can be compactly written as the following minimization
problem:

\begin{equation}
\begin{array}{ll}
\min\limits _{\textbf{H},\mathbf{J}\in\mathcal{D}}\textrm{Tr}((\textbf{H}^{T}\textbf{R}\textbf{H}){}^{-1}) & \;\textrm{s.t. }\mathbf{H}=\begin{bmatrix}\mathbf{J}\\
\mathbf{J}\mathbf{U}
\end{bmatrix}.\end{array}\label{eq:21-2}
\end{equation}
The optimization problem (\ref{eq:21-2}) has a non-convex objective
function and a non-convex constraint set. Moreover, the presence of
inverse of the quadratic term inside $\mathrm{Tr}\left(\cdot\right)$
makes (\ref{eq:21-2}) a challenging optimization problem to solve.
In the next section, we will propose an MM based algorithm to solve
this non-linear optimization problem. 

\section{PROPOSED MM BASED ALGORITHM}

In this section, we first briefly discuss the MM technique, then give
a detailed explanation of the proposed algorithm for the A optimal
design, and later discuss the computational complexity and the convergence
of the proposed algorithm. Finally, we present the extension of the
proposed algorithm for D and E optimal design. 

\subsection{MM technique}

Majorization-minimization (MM) technique is used to develop problem-driven
algorithms by exploiting the problem structure, and it works in two
steps: a majorization step and a minimization step. In the majorization
step, a surrogate function $g\left(\mathbf{x}|\mathbf{x}_{t}\right)$
that globally upperbounds the objective function $f\left(\mathbf{x}\right)$
(that is to be minimized) at a point $\mathbf{x}_{t}$ and satisfies
the following properties is constructed:
\begin{align}
g\left(\mathbf{x}|\mathbf{x}_{t}\right)\geq f\left(\mathbf{x}\right)\;\mathrm{and}\; & g\left(\mathbf{x}_{t}|\mathbf{x}_{t}\right)=f\left(\mathbf{x}_{t}\right).
\end{align}
In the minimization step, the surrogate function is minimized to give
the next update:
\begin{equation}
\mathbf{x}_{t+1}=\mathrm{arg}\;\underset{\mathbf{x}}{\mathrm{min}}\;g\left(\mathbf{x}|\mathbf{x}_{t}\right)
\end{equation}
This leads to an iterative algorithm which is repeated until convergence
to generate a non-increasing sequence $\left\{ f\left(\mathbf{x}_{t}\right)\right\} _{t\in\mathbb{N}}$,
since 
\begin{equation}
f\left(\mathbf{x}_{t+1}\right)\leq g\left(\mathbf{x}_{t+1}|\mathbf{x}_{t}\right)\leq g\left(\mathbf{x}_{t}|\mathbf{x}_{t}\right)=f\left(\mathbf{x}_{t}\right).\label{eq:23-1}
\end{equation}
The idea of MM can also be extended to maximization problems by replacing
the upperbound minimization step to lowerbound maximization step.
For more details on MM, its applications, and the construction of
surrogate functions, please refer to \cite{key-47} and the references
therein. We would like to point out here that the general description
on the MM technique included above is for problems having only primal
variable. However, when we develop our algorithm later, we will show
how the idea of MM can be extended to handle saddle point problems,
which requires constructing MM upperbounds on both primal and dual
variables.

\subsection{Proposed algorithm for A-optimal design problem}

Let us restate the optimization problem for hybrid TOA-RSS-AOA:

\begin{equation}
\begin{array}{ll}
\min\limits _{\textbf{H},\textbf{J}\in\mathcal{D}}\;\textrm{Tr}((\textbf{H}^{T}\textbf{R}\textbf{H})^{-1}) & \;\textrm{s.t. }\textbf{H}=\begin{bmatrix}\textbf{J}\\
\textbf{J}\textbf{U}
\end{bmatrix}.\end{array}\label{eqn1h}
\end{equation}
To deal with the complicated objective function in (\ref{eqn1h}),
we introduce an auxiliary variable $\mathbf{\Phi}$ and reformulate
the original problem (\ref{eqn1h}) using the Fenchel conjugate representation
\cite{key-48} of the objective function into the following equivalent
saddle-point problem: 
\begin{equation}
\begin{array}{ll}
\max\limits _{\textbf{H},\textbf{J}\in\mathcal{D}}~\min\limits _{\mathbf{\Phi}\succeq0}~\textrm{Tr}[\mathbf{\Phi}\textbf{H}^{T}\textbf{R}\textbf{H}]-\textrm{Tr}(\sqrt{\mathbf{\Phi}}) & \;\textrm{s.t. }\textbf{H}=\begin{bmatrix}\textbf{J}\\
\textbf{J}\textbf{U}
\end{bmatrix}.\end{array}\label{eqn2h}
\end{equation}
Here, the variables $\textbf{H}$ and $\textbf{J}$ can be interpreted
as the primal variables, and the variable $\mathbf{\Phi}$ as the
dual variable of the optimization problem. The equivalence of (\ref{eqn1h})
and (\ref{eqn2h}) is stated in the following Lemma.
\begin{lem}
\label{lem:1}The saddle-point problem (\ref{eqn2h}) is equivalent
to the original optimization problem (\ref{eqn1h}).
\end{lem}
\begin{IEEEproof}
The proof of the Lemma can be found in the Appendix. 
\end{IEEEproof}
We now proceed to derive an algorithm to solve the maximin problem
in (\ref{eqn2h}). Let
\begin{equation}
f\left(\textbf{H}\right)=\min\limits _{\mathbf{\Phi}\succeq0}~\textrm{Tr}[\mathbf{\Phi}\textbf{H}^{T}\textbf{R}\textbf{H}]-\textrm{Tr}(\sqrt{\mathbf{\Phi}})\label{eq:32}
\end{equation}
be a function only in the primal variable $\mathbf{H}$ and the optimization
problem in (\ref{eqn2h}) be compactly written as:
\begin{equation}
\begin{array}{ll}
\max\limits _{\textbf{H},\textbf{J}\in\mathcal{D}}~f\left(\textbf{H}\right) & \;\textrm{s.t. }\textbf{H}=\begin{bmatrix}\textbf{J}\\
\textbf{J}\textbf{U}
\end{bmatrix}.\\
\\
\end{array}\label{eq:34}
\end{equation}
For a fixed $\mathbf{\Phi}$ and $\textbf{R}\in\mathbf{S}_{+}^{m}$,
the term $\textrm{Tr}[\mathbf{\Phi}\textbf{H}^{T}\textbf{R}\textbf{H}]$
in (\ref{eq:32}) is a convex function in $\textbf{H}$. Since (\ref{eq:34})
is a maximization problem over $\mathbf{H}$, the MM technique can
be utilized to lowerbound $f\left(\textbf{H}\right)$ (at a given
$\mathbf{H}_{t}$, which of course can be obtained using a given $\mathbf{J}_{t}$)
and iteratively maximize the lowerbound. We obtain the lowerbound
via the following Lemma.
\begin{lem}
\label{lem:4-1}Given $\mathbf{H}_{t}$, the convex function $\mathrm{Tr}\left[\boldsymbol{\Phi}\mathbf{H}^{T}\mathbf{R}\mathbf{H}\right]$
can be lowerbounded as:
\begin{equation}
\mathrm{Tr}\left[\boldsymbol{\Phi}\mathbf{H}^{T}\mathbf{R}\mathbf{H}\right]\geq2\mathrm{Tr}\left[\boldsymbol{\Phi}\mathbf{H}_{t}^{T}\mathbf{R}\mathbf{H}\right]-\mathrm{Tr}\left[\boldsymbol{\Phi}\mathbf{H}_{t}^{T}\mathbf{R}\mathbf{H}_{t}\right]\label{eq:10}
\end{equation}
with equality achieved at $\mathbf{H}=\mathbf{H}_{t}$.
\end{lem}
\begin{IEEEproof}
The proof is based on the details in Section III-A of \cite{key-47}. 
\end{IEEEproof}
Using (\ref{eq:10}), $f\left(\textbf{H}\right)$ can be lowerbounded
as:
\begin{equation}
\begin{aligned}f\left(\textbf{H}\right) & \geq\min\limits _{\mathbf{\Phi}\succeq0}\;2\textrm{Tr}[\mathbf{\Phi}\textbf{H}_{t}^{T}\textbf{R}\textbf{H}]-\textrm{Tr}[\mathbf{\Phi}\textbf{H}_{t}^{T}\textbf{R}\textbf{H}_{t}]-\textrm{Tr}(\sqrt{\mathbf{\Phi}})\\
 & \,\,\,\,\,\,\,\,\,\,\,\,\,\,\,\,\,\,\,\,\,\,\,\,\,\,\,\,\,\,\,\,\,\,\,\,\,\,\,\,\,\,\,\,\,\,\,\,\,\,\,\,\,\,\:\,\,\,\,\,\,\,\,\,\,\,\,\,\,\,\,\,\,\,\,\,\,\,\,\,\,\,\,\,\,\,\,\,\,\triangleq g_{f}\left(\mathbf{H}|\textbf{\ensuremath{\mathbf{H}}}_{t}\right)
\end{aligned}
\end{equation}
giving rise to the following surrogate maximization problem:
\begin{equation}
\begin{aligned}\underset{\mathbf{H},\mathbf{J}\in\mathcal{D}}{\mathrm{max}} & \;\underset{\mathbf{\Phi}\succeq0}{\mathrm{min}}~2\textrm{Tr}[\mathbf{\Phi}\mathbf{H}_{t}^{T}\mathbf{R}\mathbf{H}]-\textrm{Tr}[\mathbf{\Phi}\textbf{\ensuremath{\mathbf{H}}}_{t}^{T}\mathbf{R}\mathbf{H}_{t}]-\textrm{Tr}(\sqrt{\mathbf{\Phi}})\\
\mathrm{s.t.\,\,\,\,\,} & \;\mathbf{H}=\begin{bmatrix}\mathbf{J}\\
\mathbf{J}\mathbf{U}
\end{bmatrix}
\end{aligned}
\label{eqn3h1}
\end{equation}
On substituting for $\textbf{R}$ and $\textbf{H}$ in \eqref{eqn3h1},
we get the following optimization problem: 
\begin{equation}
\begin{aligned}\underset{\mathbf{J}\in\mathcal{D}}{\mathrm{max}}\;\underset{\mathbf{\Phi}\succeq0}{\mathrm{min}}\; & 2\textrm{Tr}[\mathbf{\Phi}\mathbf{J}_{t}^{T}\mathbf{R}_{1}\mathbf{J}]+2\textrm{Tr}[\mathbf{\Phi}\mathbf{U}^{T}\mathrm{\mathbf{J}}_{t}^{T}\mathbf{R}_{2}\mathbf{J}\mathbf{U}]-\textrm{Tr}(\sqrt{\mathbf{\Phi}})\\
 & -\textrm{Tr}[\mathbf{\Phi}\mathbf{J}_{t}^{T}\mathbf{R}_{1}\mathbf{J}_{t}]-\textrm{Tr}[\mathbf{\Phi}\mathbf{U}^{T}\mathbf{J}_{t}^{T}\mathbf{R}_{2}\mathbf{J}_{t}\mathbf{U}].
\end{aligned}
\label{eqn3h}
\end{equation}
The constraint set $\mathcal{D}$ for $\mathbf{J}$ can be relaxed
to $\mathcal{D}_{r}$ (which is defined as $\left\{ \mathbf{j}_{i}\right\} $'s
satisfying $\textbf{j}_{i}^{T}\textbf{j}_{i}\leq1~\forall~i$) as
the objective in (\ref{eqn3h}) is linear in $\mathbf{J}$, and the
optimum over $\mathbf{J}$ will occur only on the boundary and hence
the relaxation will always be tight. With this relaxation, we can
now reformulate the maximin problem in (\ref{eqn3h}) into a minimax
problem using the following minimax theorem:
\begin{thm}
\cite{key-49}\label{thm:3} Let $\mathcal{X}$ and $\mathcal{Y}$
be compact convex sets. If $f:\mathcal{X}\times\mathcal{Y}\rightarrow\mathbb{R}$
is a continuous function that is concave-convex, i.e., $f\left(\cdot,y\right):\mathcal{X\rightarrow\mathbb{R}}$
is concave for fixed $y$, and $f\left(y,\cdot\right):\mathcal{Y\rightarrow\mathbb{R}}$
is convex for fixed $x$, then we have
\begin{equation}
\underset{x\in\mathcal{X}}{\mathrm{max}}\;\underset{y\in\mathcal{Y}}{\mathrm{min}}\;f\left(x,y\right)=\underset{y\in\mathcal{Y}}{\mathrm{min}}\;\underset{x\in\mathcal{X}}{\mathrm{max}}\;f\left(x,y\right)
\end{equation}
\end{thm}
Since the constraint sets (after relaxation) are compact convex sets,
and the objective function in (\ref{eqn3h}) is continuous, and convex
in $\mathbf{\Phi}$ for a fixed $\mathbf{J}$ (the term $-\textrm{Tr}(\sqrt{\mathbf{\Phi}})$
is a convex function in $\mathbf{\Phi}$ via the Lieb's concavity
theorem \cite{key-50}) and linear in $\mathbf{J}$ for a fixed $\mathbf{\Phi}$,
we can switch the max and min term using Theorem \ref{thm:3} to arrive
at the following minimax problem:
\begin{equation}
\begin{aligned}\min\limits _{\mathbf{\Phi}\succeq0}~\max\limits _{\textbf{J}\in\mathcal{D}_{r}}~ & 2\textrm{Tr}[\mathbf{\Phi}\textbf{J}_{t}^{T}\textbf{R}_{1}\textbf{J}]+2\textrm{Tr}[\mathbf{\Phi}\textbf{U}^{T}\textbf{J}_{t}^{T}\textbf{R}_{2}\textbf{J}\textbf{U}]-\textrm{Tr}(\sqrt{\mathbf{\Phi}})\\
 & -\textrm{Tr}[\mathbf{\Phi}\textbf{J}_{t}^{T}\textbf{R}_{1}\textbf{J}_{t}]-\textrm{Tr}[\mathbf{\Phi}\textbf{U}^{T}\textbf{J}_{t}^{T}\textbf{R}_{2}\textbf{J}_{t}\textbf{U}]
\end{aligned}
\label{eqn4h}
\end{equation}
The minimax problem (\ref{eqn4h}) has an inner maximization problem
in $\mathbf{J}$ and an outer minimization problem in $\mathbf{\Phi}$.
We will first solve the inner maximization problem in \eqref{eqn4h}.
Using $\textbf{Q}\triangleq\textbf{U}\mathbf{\Phi}\textbf{U}^{T}$,
the inner maximization problem be written as: 
\begin{equation}
\begin{array}{ll}
\max\limits _{\textbf{J}\in D_{r}}\;2\textrm{Tr}[\textbf{J}_{t}^{T}\textbf{R}_{1}\textbf{J}\mathbf{\Phi}]+2\textrm{Tr}[\textbf{J}_{t}^{T}\textbf{R}_{2}\textbf{J}\textbf{Q}],\end{array}\label{eq:20}
\end{equation}
Let 
\begin{equation}
\mathbf{Z}=\mathbf{\Phi}\tilde{\mathbf{A}}_{1}+\textbf{Q}\tilde{\mathbf{A}}_{2},\label{eq:22}
\end{equation}
with $\tilde{\mathbf{A}}_{1}\triangleq\textbf{J}_{t}^{T}\textbf{R}_{1}$
and $\tilde{\mathbf{A}}_{2}\triangleq\textbf{J}_{t}^{T}\textbf{R}_{2}$,
then the optimization problem in (\ref{eq:20}) can be rewritten as:
\begin{equation}
\max\limits _{\textbf{J}\in D_{r}}\;2\sum_{i=1}^{m}\mathbf{j}_{i}^{T}\mathbf{z}_{i},\label{eq:23}
\end{equation}
where $\mathbf{z}_{i}$ denotes the $i^{\mathrm{th}}$ column $\mathbf{Z}$.
Problem (\ref{eq:23}) has the following closed-form solution: 
\begin{equation}
\textbf{j}_{i}^{*}=\frac{\mathbf{z}_{i}}{\|\mathbf{z}_{i}\|}\label{eq:24}
\end{equation}
Please note that the optimum occurs on the boundary of the constraint
set $\mathcal{D}_{r}$ as noted before.

Eliminating the variable $\mathbf{J}$ from (\ref{eqn4h}) by substituting
for the optimal minimizer over $\mathbf{J}$ (i.e., $\textbf{j}_{i}^{*}=\frac{\mathbf{z}_{i}}{\|\mathbf{z}_{i}\|}$)
gives the following minimization problem in the dual variable $\mathbf{\Phi}$:
\begin{equation}
\begin{array}{ll}
\min\limits _{\mathbf{\Phi}\succeq0} & 2\stackrel[i=1]{m}{\sum}||\mathbf{z}_{i}||-\textrm{Tr}[\mathbf{\Phi}\textbf{J}_{t}^{T}\textbf{R}_{1}\textbf{J}_{t}]-\textrm{Tr}[\mathbf{\Phi}\textbf{U}^{T}\textbf{J}_{t}^{T}\textbf{R}_{2}\textbf{J}_{t}\textbf{U}]\\
 & -\textrm{Tr}(\sqrt{\mathbf{\Phi}}).
\end{array}\label{eqn6h}
\end{equation}
Substituting for $\mathbf{z}_{i}$ from (\ref{eq:22}) in (\ref{eqn6h}),
we get:
\begin{equation}
\begin{array}{ll}
\min\limits _{\mathbf{\Phi}\succeq0} & 2\stackrel[i=1]{m}{\sum}\|\mathbf{\Phi}\mathbf{\tilde{\mathbf{a}}}_{i}^{1}+\textbf{Q}\tilde{\mathbf{a}}_{i}^{2}\|-\textrm{Tr}[\textbf{J}_{t}^{T}\textbf{R}_{1}\textbf{J}_{t}\mathbf{\Phi}]\\
 & -\textrm{Tr}[\textbf{U}^{T}\textbf{J}_{t}^{T}\textbf{R}_{2}\textbf{J}_{t}\textbf{U}\mathbf{\Phi}]-\textrm{Tr}(\sqrt{\mathbf{\Phi}}),
\end{array}\label{eq:49}
\end{equation}
where $\tilde{\mathbf{a}}_{i}^{1}$ and $\tilde{\mathbf{a}}_{i}^{2}$
denote the $i^{\mathrm{th}}$ column of $\tilde{\mathbf{A}}_{1}$
and $\tilde{\mathbf{A}}_{2}$, respectively. Problem in (\ref{eq:49})
can be rewritten as:
\begin{equation}
\begin{array}{ll}
\min\limits _{\mathbf{\Phi}\succeq0} & 2\stackrel[i=1]{m}{\sum}\left\Vert \begin{bmatrix}\mathbf{\Phi},\textbf{Q}\end{bmatrix}\begin{bmatrix}\tilde{\mathbf{a}}_{i}^{1}\\
\tilde{\mathbf{a}}_{i}^{2}
\end{bmatrix}\right\Vert -\textrm{Tr}[\textbf{J}_{t}^{T}\textbf{R}_{1}\textbf{J}_{t}\mathbf{\Phi}]\\
 & -\textrm{Tr}[\textbf{U}^{T}\textbf{J}_{t}^{T}\textbf{R}_{2}\textbf{J}_{t}\textbf{U}\mathbf{\Phi}]-\textrm{Tr}(\sqrt{\mathbf{\Phi}}).
\end{array}\label{eq:27}
\end{equation}
Using $\tilde{\textbf{\ensuremath{\mathbf{\Phi}}}}=\begin{bmatrix}\mathbf{\Phi},\textbf{Q}\end{bmatrix}$
and $\mathbf{a}_{i}=\begin{bmatrix}\tilde{\mathbf{a}}_{i}^{1}\\
\tilde{\mathbf{a}}_{i}^{2}
\end{bmatrix}$, the optimization problem (\ref{eq:27}) can be compactly written
as:
\begin{equation}
\min\limits _{\mathbf{\Phi}\succeq0}~2\stackrel[i=1]{m}{\sum}\|\tilde{\textbf{\ensuremath{\mathbf{\Phi}}}}\mathbf{a}_{i}\|-\textrm{Tr}[\textbf{H}_{t}^{T}\textbf{R}\textbf{H}_{t}\mathbf{\Phi}]-\textrm{Tr}(\sqrt{\mathbf{\Phi}})\triangleq h\left(\mathbf{\Phi}\right).\label{eq:28}
\end{equation}
The minimization problem (\ref{eq:28}) is a convex problem (semidefinite
programming problem (SDP)) in $\mathbf{\Phi}$ and can be solved using
some off-the-shelf interior point solvers. However, solving such SDP
via the standard solvers will be computationally inefficient and will
limit the application of our approach to only low dimensional problems.
Therefore, we once again use ``MM'' to solve (\ref{eq:28}) which
results in a double loop MM algorithm (by double loop we mean loop
inside a loop). To this end, we construct a surrogate for the objective
in (\ref{eq:28}) at some $\mathbf{\Phi}=\mathbf{\Phi}_{k}$. Since,
(\ref{eq:28}) is a minimization problem, we first upperbound the
term $\|\tilde{\textbf{\ensuremath{\mathbf{\Phi}}}}\mathbf{a}_{i}\|$
in $h\left(\mathbf{\Phi}\right)$ as per the following Lemma: 
\begin{lem}
\label{lem:4}Given $\tilde{\mathbf{\Phi}}_{k}$, the $\ell_{2}$
norm function $\|\tilde{\textbf{\ensuremath{\mathbf{\Phi}}}}\mathbf{a}_{i}\|$
can be upperbounded as:
\begin{equation}
\|\tilde{\textbf{\ensuremath{\mathbf{\Phi}}}}\mathbf{a}_{i}\|\leq\frac{\|\tilde{\textbf{\ensuremath{\mathbf{\Phi}}}}\mathbf{a}_{i}\|^{2}}{2\|\tilde{\textbf{\ensuremath{\mathbf{\Phi}}}}_{k}\mathbf{a}_{i}\|}+\mathrm{const}.,\label{29}
\end{equation}
given that $||\tilde{\textbf{\ensuremath{\mathbf{\Phi}}}}_{k}\mathbf{a}_{i}||\neq0$.
Equality is achieved at $\mathbf{\Phi}=\mathbf{\Phi}_{k}$.
\end{lem}
\begin{IEEEproof}
The proof can be established using the details in Section III-A of
\cite{key-47}.
\end{IEEEproof}
Using Lemma \ref{lem:4}, we can upperbound $h\left(\mathbf{\Phi}\right)$
as:\\
\begin{equation}
\begin{array}{ll}
h\left(\mathbf{\Phi}\right)\leq & \stackrel[i=1]{m}{\sum}\frac{\|\tilde{\textbf{\ensuremath{\mathbf{\Phi}}}}\mathbf{a}_{i}\|^{2}}{\|\tilde{\textbf{\ensuremath{\mathbf{\Phi}}}}_{k}\mathbf{a}_{i}\|}-\textrm{Tr}[\textbf{H}_{t}^{T}\textbf{R}\textbf{H}_{t}\mathbf{\Phi}]-\textrm{Tr}(\sqrt{\mathbf{\Phi}}),\end{array}\label{eqn8h}
\end{equation}
which can also be written as:\\
\begin{align}
h\left(\mathbf{\Phi}\right)\leq & \stackrel[i=1]{m}{\sum}\frac{\mathbf{a}_{i}^{T}\tilde{\textbf{\ensuremath{\mathbf{\Phi}}}}^{T}\tilde{\textbf{\ensuremath{\mathbf{\Phi}}}}\mathbf{a}_{i}}{\|\tilde{\textbf{\ensuremath{\mathbf{\Phi}}}}_{k}\mathbf{a}_{i}\|}-\textrm{Tr}[\textbf{H}_{t}^{T}\textbf{R}\textbf{H}_{t}\mathbf{\Phi}]-\textrm{Tr}(\sqrt{\mathbf{\Phi}}).\label{eq:48}
\end{align}
Using $\mathbf{A}=\stackrel[i=1]{m}{\sum}\frac{\mathbf{a}_{i}\mathbf{a}_{i}^{T}}{\|\tilde{\textbf{\ensuremath{\mathbf{\Phi}}}}_{k}\mathbf{a}_{i}\|}$,
(\ref{eq:48}) can be rewritten as:
\begin{equation}
\begin{array}{ll}
h\left(\mathbf{\Phi}\right)\leq & \textrm{Tr}(\tilde{\mathbf{\Phi}}\mathbf{A}\tilde{\mathbf{\Phi}}^{T})-\textrm{Tr}[\mathbf{\Phi}\textbf{H}_{t}^{T}\textbf{R}\textbf{H}_{t}]-\textrm{Tr}(\sqrt{\mathbf{\Phi}})\\
 & \,\,\,\,\,\,\,\,\,\,\,\,\,\,\,\,\,\,\,\,\,\,\,\,\,\,\,\,\,\,\,\,\,\,\,\,\,\,\,\,\,\,\:\,\,\,\,\,\,\,\,\,\,\,\,\,\,\,\,\,\,\,\,\,\,\,\,\,\,\,\,\,\,\,\,\,\,\,\triangleq\tilde{g}_{h}\left(\mathbf{\Phi}|\mathbf{\Phi}_{k}\right)
\end{array}\label{eqn9h}
\end{equation}
and the associated surrogate minimization problem can be written as:
\begin{equation}
\min\limits _{\mathbf{\Phi}\succeq0}~\tilde{g}_{h}\left(\mathbf{\Phi}|\mathbf{\Phi}_{k}\right).\label{eq:50}
\end{equation}
Problem (\ref{eq:50}) does not have a closed-form solution and therefore,
we once again upperbound $\tilde{g}_{h}\left(\mathbf{\Phi}|\mathbf{\Phi}_{k}\right)$
at $\mathbf{\Phi}=\mathbf{\Phi}_{k}$. Let $\lambda_{1}=\lambda_{\mathrm{max}}\left(\mathbf{A}\right)$
and $\tilde{\mathbf{A}}=\mathbf{A}-\lambda_{1}\mathbf{I}$, then (\ref{eq:50})
can be rewritten as:
\begin{equation}
\min\limits _{\mathbf{\Phi}\succeq0}~\textrm{Tr}(\tilde{\mathbf{\Phi}}\tilde{\mathbf{A}}\tilde{\mathbf{\Phi}}^{T})+\lambda_{1}\textrm{Tr}\left(\tilde{\mathbf{\Phi}}^{T}\tilde{\mathbf{\Phi}}\right)-\textrm{Tr}[\textbf{H}_{t}^{T}\textbf{R}\textbf{H}_{t}\mathbf{\Phi}]-\textrm{Tr}(\sqrt{\mathbf{\Phi}}).\label{eq:51}
\end{equation}
The first term $\textrm{Tr}(\tilde{\mathbf{\Phi}}\tilde{\mathbf{A}}\tilde{\mathbf{\Phi}}^{T})$
is a concave function in $\tilde{\mathbf{\Phi}}$ and can be linearized
using the first order Taylor series expansion as:
\begin{equation}
\textrm{Tr}(\tilde{\mathbf{\Phi}}\tilde{\mathbf{A}}\tilde{\mathbf{\Phi}}^{T})\leq-\textrm{Tr}(\tilde{\mathbf{\Phi}}_{k}\tilde{\mathbf{A}}\tilde{\mathbf{\Phi}}_{k}^{T})+2\textrm{Tr}(\tilde{\mathbf{\Phi}}_{k}\tilde{\mathbf{A}}\tilde{\mathbf{\Phi}}^{T})\label{eq:52}
\end{equation}
Using (\ref{eq:52}), $\tilde{g}_{h}\left(\mathbf{\Phi}|\mathbf{\Phi}_{k}\right)$
can be upperbounded as:

\begin{equation}
\begin{aligned}\tilde{g}_{h}\left(\mathbf{\Phi}|\mathbf{\Phi}_{k}\right)\leq & -\textrm{Tr}(\tilde{\mathbf{\Phi}}_{k}\tilde{\mathbf{A}}\tilde{\mathbf{\Phi}}_{k}^{T})+2\textrm{Tr}(\tilde{\mathbf{\Phi}}_{k}\tilde{\mathbf{A}}\tilde{\mathbf{\Phi}}^{T})+\lambda_{1}\textrm{Tr}\left(\tilde{\mathbf{\Phi}}^{T}\tilde{\mathbf{\Phi}}\right)\\
 & -\textrm{Tr}[\textbf{H}_{t}^{T}\textbf{R}\textbf{H}_{t}\mathbf{\Phi}]-\textrm{Tr}(\sqrt{\mathbf{\Phi}})\triangleq g_{h}\left(\mathbf{\Phi}|\mathbf{\Phi}_{k}\right).
\end{aligned}
\end{equation}
It is to be noted that $g_{h}\left(\mathbf{\Phi}|\mathbf{\Phi}_{k}\right)$
will also tightly upperbound $h\left(\mathbf{\Phi}\right)$ at $\mathbf{\Phi}=\mathbf{\Phi}_{k}$.
Thus, we arrive at the following surrogate minimization problem:
\begin{equation}
\begin{aligned}\min\limits _{\mathbf{\Phi}\succeq0}\; & 2\textrm{Tr}(\tilde{\mathbf{\Phi}}_{k}\tilde{\mathbf{A}}\tilde{\mathbf{\Phi}}^{T})+\lambda_{1}\textrm{Tr}\left(\tilde{\mathbf{\Phi}}^{T}\tilde{\mathbf{\Phi}}\right)-\textrm{Tr}[\textbf{H}_{t}^{T}\textbf{R}\textbf{H}_{t}\mathbf{\Phi}]\\
 & -\textrm{Tr}(\sqrt{\mathbf{\Phi}}).
\end{aligned}
\label{eqn10h}
\end{equation}
Substituting for $\tilde{\mathbf{\Phi}}$ in (\ref{eqn10h}) and using
$\mathbf{B}=\left[\mathbf{B}_{1}\boldsymbol{,}\mathbf{B}_{2}\right]=\tilde{\mathbf{\Phi}}_{k}\tilde{\mathbf{A}}$,
we get the following optimization problem: 

\begin{equation}
\begin{array}{ll}
\min\limits _{\mathbf{\Phi}\succeq0}\; & 2\textrm{Tr}\left(\left[\mathbf{B}_{1}\boldsymbol{,}\mathbf{B}_{2}\right]\begin{bmatrix}\mathbf{\Phi}\\
\mathbf{Q}
\end{bmatrix}\right)+2\lambda_{1}\textrm{Tr}\left(\mathbf{\Phi}^{2}\right)-\textrm{Tr}[\textbf{H}_{t}^{T}\textbf{R}\textbf{H}_{t}\mathbf{\Phi}]\\
 & -\textrm{Tr}(\sqrt{\mathbf{\Phi}}).
\end{array}\label{eqn10h1}
\end{equation}
With $\lambda=2\lambda_{1}$ and substituting $\mathbf{Q}$ in (\ref{eqn10h1}),
we get:
\begin{equation}
\begin{aligned}\min\limits _{\mathbf{\Phi}\succeq0}\; & \textrm{Tr}((\mathbf{B}_{1}+\mathbf{B}_{1}^{T})\mathbf{\Phi})+\textrm{Tr}(\mathbf{U}^{T}(\mathbf{B}_{2}+\mathbf{B}_{2}^{T})\mathbf{U}\mathbf{\mathbf{\Phi}})+\lambda\textrm{Tr}\left(\mathbf{\Phi}^{2}\right)\\
 & -\textrm{Tr}[\textbf{H}_{t}^{T}\textbf{R}\textbf{H}_{t}\mathbf{\Phi}]-\textrm{Tr}(\sqrt{\mathbf{\Phi}})
\end{aligned}
\label{47}
\end{equation}
Using $\textbf{C}=\mathbf{B}_{1}+\mathbf{B}_{1}^{T}+\mathbf{U}^{T}(\mathbf{B}_{2}+\mathbf{B}_{2}^{T})\mathbf{U}-\textbf{H}_{t}^{T}\textbf{R}\textbf{H}_{t}$,
(\ref{47}) can be compactly written as:
\begin{equation}
\min\limits _{\mathbf{\Phi}\succeq0}\;\textrm{Tr}(\textbf{C}\mathbf{\Phi})+\lambda\textrm{Tr}(\boldsymbol{\mathbf{\Phi}^{2}})-\textrm{Tr}(\sqrt{\mathbf{\Phi}})
\end{equation}
The optimal solution of the above mentioned problem can be computed
easily based on the eigenvalue decomposition of $\mathbf{C}$ and
solving scalar cubic equations. Suppose $\mathbf{V}$ and $\mathbf{E}$
denote the matrices containing the eigenvectors and eigenvalues of
$\mathbf{C}$, then the minimizer for the problem will be $\mathbf{V}\mathbf{X}\mathbf{V}^{T}$,
where any diagonal element of the matrix $\mathbf{X}$ will be computed
by solving the following scalar problems: 
\begin{equation}
\begin{array}{ll}
\min\limits _{X_{ii}>0}E_{ii}X_{ii}+\lambda X_{ii}^{2}-\sqrt{X_{ii}} & \;\forall i\end{array}\label{eqn13h}
\end{equation}
The KKT condition for the above problem will be: 
\begin{equation}
\begin{array}{ll}
E_{ii}+2\lambda X_{ii}^{*}-\frac{1}{2\sqrt{X_{ii}^{*}}}=0 & \;\forall i\end{array}\label{eqn14h}
\end{equation}
Using $\tilde{X}_{ii}^{*}=\sqrt{X_{ii}^{*}}$, and then multiplying
both sides of equation (\ref{eqn14h}) with $2\tilde{X}_{ii}^{*}$
gives: 
\begin{equation}
\begin{array}{ll}
4\lambda\left(\tilde{X}{}_{ii}^{*}\right)^{3}+2E_{ii}\tilde{X}_{ii}^{*}-1=0 & \;\forall i\end{array},\label{eqn16h}
\end{equation}
which is nothing but a cubic equation whose positive real root will
give the optimum $\mathbf{X}^{*}$.

\subsection{Solving the D and E optimal design problems}

The algorithm proposed in the previous subsection can be extended
to D and E optimal criteria as follows. 

In D-optimal design, the problem is to solve the following maximization
problem:
\begin{equation}
\begin{aligned}\underset{\mathbf{H},\mathbf{J}\in\mathcal{D}}{\mathrm{max}} & \;\mathrm{log}\left|\mathbf{H}^{T}\mathbf{R}\mathbf{H}\right|,\;\textrm{s.t. \textbf{H}}=\begin{bmatrix}\textbf{J}\\
\textbf{J}\textbf{U}
\end{bmatrix}.\end{aligned}
\label{eq:1}
\end{equation}
The presence of quadratic term inside $\mathrm{log}\left|\cdot\right|$
makes (\ref{eq:1}) a challenging problem to solve. Similar to the
A-optimal case, we use the Fenchel conjugate representation of the
log determinant function, which is given as:
\begin{equation}
\mathrm{log}\left|\mathbf{A}\right|=\underset{\mathbf{B}\succeq0}{\mathrm{min}}\:\mathbf{Tr}\left(\mathbf{A}\mathbf{B}\right)-\mathrm{log}\left|\mathbf{B}\right|,\label{eq:2}
\end{equation}
to reformulate problem (\ref{eq:1}) into the following equivalent
saddle-point problem: 
\begin{equation}
\begin{aligned}\max\limits _{\textbf{H},\mathbf{J}\in\mathcal{D}} & \;\min\limits _{\mathbf{\Phi}\succeq0}~\textrm{Tr}[\mathbf{\Phi}\textbf{H}^{T}\textbf{R}\textbf{H}]-\log|\mathbf{\Phi}|\\
\textrm{s. t. \,\,\,} & \textbf{H}=\begin{bmatrix}\textbf{J}\\
\textbf{J}\textbf{U}
\end{bmatrix}.
\end{aligned}
\label{eq:3-1}
\end{equation}
Similar to the proof of Lemma \ref{lem:1} in the A-optimal design,
it can be shown that the maximin problem (\ref{eq:3-1}) is equivalent
to the problem (\ref{eq:1}). 

In the case of E-optimal design, the objective is to solve the following
maximization problem: 
\begin{equation}
\begin{aligned}\underset{\textbf{H},\mathbf{J}\in\mathcal{D}}{\mathrm{max}}\; & \mathrm{\lambda}_{\mathrm{min}}\left(\mathbf{H}^{T}\mathbf{R}\mathbf{H}\right)\;\textrm{s.t. \textbf{H}}=\begin{bmatrix}\textbf{J}\\
\textbf{J}\textbf{U}
\end{bmatrix}.\end{aligned}
\label{eq:57}
\end{equation}
(Typically, in E optimal design, the objective is to minimize the
maximum eigen value of the matrix $\left(\mathbf{H}^{T}\mathbf{R}\mathbf{H}\right)^{-1}$
which is equivalent to maximizing the minimum eigen value of the matrix
$\left(\mathbf{H}^{T}\mathbf{R}\mathbf{H}\right)$). Unlike A and
D optimal case, there is no such Fenchel representation readily available
for the $\mathrm{\lambda}_{\mathrm{min}}\left(\cdot\right)$ function.
So, we resort to the Lagrangian dual formulation to arrive at the
saddle-point formulation. We first consider the epigraph reformulation
of the problem in (\ref{eq:57}):
\begin{equation}
\begin{aligned}\underset{\mathbf{H},\mathbf{\mathbf{J}\in\mathcal{D}},\alpha}{\mathrm{max}} & \;\alpha\\
\textrm{s.t. }\,\,\,\,\,\, & \;\textbf{H}^{T}\textbf{R}\textbf{H}\succeq\alpha\mathbf{I},\;\textbf{H}=\begin{bmatrix}\textbf{J}\\
\textbf{J}\textbf{U}
\end{bmatrix}.
\end{aligned}
\label{eq:58}
\end{equation}
The Lagrangian dual of the problem in (\ref{eq:58}) is given by:
\begin{equation}
\begin{aligned}\max\limits _{\textbf{H},\mathbf{J}\in\mathcal{D},\alpha}\:\min\limits _{\mathbf{\Phi}\succeq0} & ~\alpha+\textrm{Tr}[\mathbf{\Phi}\textbf{H}^{T}\textbf{R}\textbf{H}]-\alpha\textrm{Tr}[\mathbf{\Phi}]\\
\textrm{s.t. }\,\,\,\,\,\,\,\,\,\,\,\,\,\,\,\,\,\, & \;\textbf{H}=\begin{bmatrix}\textbf{J}\\
\textbf{J}\textbf{U}
\end{bmatrix},
\end{aligned}
\label{eq:59}
\end{equation}
where $\mathbf{\Phi}$ is the Lagrangian dual variable. On rearranging
the terms in (\ref{eq:59}), we get:
\begin{equation}
\begin{aligned}\max\limits _{\textbf{H},\mathbf{J}\in\mathcal{D},\alpha}\:\min\limits _{\mathbf{\Phi}\succeq0} & ~\textrm{Tr}[\mathbf{\Phi}\textbf{H}^{T}\textbf{R}\textbf{H}]-\alpha\left(1-\textrm{Tr}[\mathbf{\Phi}]\right)\\
\textrm{s.t.\,\,\,\,\,\,\,\,\,\,\,\,\,\,\,\,\,\, } & \;\textbf{H}=\begin{bmatrix}\textbf{J}\\
\textbf{J}\textbf{U}
\end{bmatrix}.
\end{aligned}
\label{eq:60}
\end{equation}

Here we would like to note that in the saddle-point formulations in
(\ref{eqn2h}), (\ref{eq:3-1}) and (\ref{eq:60}), the term containing
the primal variable `$\mathbf{H}$' remains the same, and the difference
is in the second term (which is a function only in the dual variable
`$\mathbf{\Phi}$'). Thus, the difference between the algorithms for
the different optimal criteria will only be in the update for `$\mathbf{\Phi}$'.
Therefore, in the following we will discuss only necessary changes
in the update of $\mathbf{\Phi}$ for D and E optimal design problems.

In D-optimal design, to update $\mathbf{\Phi}$, we construct surrogate
using MM (at $\mathbf{\Phi}=\mathbf{\Phi}_{k}$) in a manner similar
to the A-optimal design and get the following minimization problems:
\vspace{-12pt}

\begin{equation}
\begin{array}{ll}
\min\limits _{X_{ii}>0}E_{ii}X_{ii}+\lambda X_{ii}^{2}-\log(X_{ii}) & \;\forall i\end{array}\label{eq13}
\end{equation}
The KKT condition for the above problem is: 
\begin{equation}
\begin{array}{ll}
E_{ii}+2\lambda X_{ii}^{*}-\frac{1}{X_{ii}^{*}} & =0\end{array}\;\forall i\label{eqkkt}
\end{equation}
which is a quadratic equation and the optimal solution will be its
positive root:
\begin{equation}
X_{ii}^{*}=\frac{-E_{ii}+\sqrt{E_{ii}^{2}+8\lambda}}{4\lambda_{}}\;\;\forall i\label{eq:63}
\end{equation}

In E-optimal design case, following the same steps as in A-optimal
design, we arrive at the following surrogate (minimax) problem: 
\begin{equation}
\begin{aligned}\min\limits _{\mathbf{\Phi}\succeq0}~\max\limits _{\textbf{J}\in\mathcal{D}_{r},\alpha}\; & 2\textrm{Tr}[\textbf{J}_{t}^{T}\textbf{R}_{1}\textbf{J}\mathbf{\Phi}]-\textrm{Tr}[\textbf{J}_{t}^{T}\textbf{R}_{1}\textbf{J}_{t}\mathbf{\Phi}]+2\textrm{Tr}[\textbf{J}_{t}^{T}\textbf{R}_{2}\textbf{J}\mathbf{Q}]\\
 & -\textrm{Tr}[\textbf{J}_{t}^{T}\textbf{R}_{2}\textbf{J}_{t}\mathbf{Q}]+\alpha(1-\textrm{Tr}[\mathbf{\Phi}]).
\end{aligned}
\end{equation}
Here, we first eliminate the variable $\alpha$ and get:
\begin{equation}
\begin{aligned}\min\limits _{\mathbf{\Phi}\succeq0}~\max\limits _{\textbf{J}\in\mathcal{D}_{r}} & \;2\textrm{Tr}[\textbf{J}_{t}^{T}\textbf{R}_{1}\textbf{J}\mathbf{\Phi}]-\textrm{Tr}[\textbf{J}_{t}^{T}\textbf{R}_{1}\textbf{J}_{t}\mathbf{\Phi}]+2\textrm{Tr}[\textbf{J}_{t}^{T}\textbf{R}_{2}\textbf{J}\mathbf{Q}]\\
 & \,\,\,\,\,\,\,\,\,\,\,\,\,\,\,\,\,\,\,\,\,\,\,\,\,\,\,\,\,\,\,\,\,\,\,\,\,\,\,\,\,\,\,\,\,\,\,\,\,\,\,\,\,\,\,\,\,\,\,\,\,\,\,\,\,\,\,\,\,\,\,\,\,\,-\textrm{Tr}[\textbf{J}_{t}^{T}\textbf{R}_{2}\textbf{J}_{t}\mathbf{Q}]\\
\textrm{s.t. \,\,\,\,\,\,\,\,\,\,\,\,\,\,\,} & \;\textrm{Tr}[\mathbf{\Phi}]\geq1.
\end{aligned}
\label{eq:67}
\end{equation}
The constraint in (\ref{eq:67}) will be tight as the objective in
(\ref{eq:67}) is linear in $\mathbf{\Phi}$, and therefore, the optimum
will occur at the boundary of the constraint set. Following the steps
similar to the A-optimal design for eliminating $\mathbf{J}$, we
arrive at the following optimization problem:
\begin{equation}
\begin{aligned}\min\limits _{\mathbf{\Phi}\succeq0} & \;2\sum_{i=1}^{m}\left\Vert \tilde{\textbf{\ensuremath{\mathbf{\Phi}}}}\mathbf{a}_{i}\right\Vert -\textrm{Tr}[\mathbf{\Phi}\textbf{H}_{t}^{T}\textbf{R}\textbf{H}_{t}]\\
\textrm{s.t. } & \;\textrm{Tr}[\mathbf{\Phi}]=1,
\end{aligned}
\label{eq:66}
\end{equation}
Problem in (\ref{eq:66}) is a convex optimization problem (an SDP).
Unlike the A and D optimal case, solving the problem in (\ref{eq:66})
via MM is very challenging, so we prefer to solve (\ref{eq:66}) via
standard interior point methods like CVX. The pseudo-code of the proposed
algorithm for TOA-RSS-AOA measurements is given in Algorithm 1, with
Algorithm 2 and Algorithm 3 describing the $\mathbf{\Phi}$ update
for the three optimal designs.

\begin{algorithm}[tbh]
\caption{Pseudo-code of the proposed algorithm (A, D and E optimal design)}

\begin{lyxlist}{00.00.0000}
\item [{\textbf{Input:}}] $m$, $n$, $\mathbf{D}$, $\boldsymbol{\Sigma}_{\text{TOA}}$,
$\boldsymbol{\Sigma}_{\text{RSS}}$, $\boldsymbol{\Sigma}_{\text{AOA}}$,
\textbf{$\boldsymbol{\mathbf{U}}$ }and $\epsilon=10^{-3}$
\begin{enumerate}
\item Initialize $\mathbf{J}_{0}$ and $\boldsymbol{\Phi}{}_{0}$
\item \textbf{Iterate:} Given $\mathbf{J}_{t}$, do the $\left(t+1\right)^{\mathbf{th}}$
step.
\begin{enumerate}
\item Compute $\mathbf{H}_{t}$, $\mathbf{\tilde{\mathbf{A}}_{1}}$ and
$\tilde{\mathbf{A}}_{2}$. 
\item Update $\boldsymbol{\Phi}$ using Algorithm 2 (for A- and D- optimal
design) and Algorithm 3 (for E-optimal design).
\item Compute $\mathbf{Z}$.
\item Solve (\ref{eq:24}) to obtain $\mathbf{J}_{t+1}$.
\item If $\left\Vert \mathbf{J}_{t+1}-\mathbf{J}_{t}\right\Vert _{F}/\left\Vert \mathbf{J}_{t}\right\Vert _{F}<\epsilon$,
stop and return $\mathbf{J}_{t+1}$.
\end{enumerate}
\end{enumerate}
\item [{\textbf{Output:}}] \textbf{$\mathbf{J}^{*}$ }is the value of $\mathbf{J}$
obtained at the convergence of the loop.
\end{lyxlist}
\end{algorithm}

\begin{algorithm}[tbh]
\caption{Pseudo code to update $\boldsymbol{\Phi}$ in A and D optimal design}

\begin{lyxlist}{00.00.0000}
\item [{\textbf{Input:}}] $\mathbf{H}_{t}$, $\tilde{\textbf{A}}_{1}$,
$\tilde{\textbf{A}}_{2}$ and $\epsilon=10^{-3}$
\end{lyxlist}
\begin{enumerate}
\item \textbf{Iterate:} Given $\boldsymbol{\Phi}{}_{k}$, do the $\left(k+1\right)^{\mathbf{th}}$
step.
\begin{enumerate}
\item Compute $\tilde{\boldsymbol{\Phi}}{}_{k}$, $\mathbf{A}$, $\lambda_{\mathrm{max}}\left(\mathbf{A}\right)$,
$\mathbf{B}=\left[\mathbf{B}_{1},\mathbf{B}_{2}\right]$ and $\mathbf{C}$.
\item Compute the EVD $\mathbf{C}=\mathbf{V}\mathbf{E}\mathbf{V}^{T}$.
\item 
\begin{lyxlist}{00.00.0000}
\item [{\textbf{If}}] A-optimal design

Solve (\ref{eqn16h}) to obtain $X_{ii}^{*}\;\forall i$.
\item [{\textbf{Elseif}}] D-optimal design

Solve (\ref{eq:63}) to obtain $X_{ii}^{*}\;\forall i$.
\end{lyxlist}
\item $\boldsymbol{\Phi}{}_{k+1}=\mathbf{V}\mathbf{X}^{*}\mathbf{V}^{T}$.
\item If $\left\Vert \mathbf{\boldsymbol{\Phi}}_{k+1}-\mathbf{\boldsymbol{\Phi}}_{k}\right\Vert _{F}/\left\Vert \mathbf{\boldsymbol{\Phi}}_{k}\right\Vert _{F}<\epsilon$,
stop and return $\mathbf{\boldsymbol{\Phi}}_{k+1}$.
\end{enumerate}
\end{enumerate}
\begin{lyxlist}{00.00.0000}
\item [{\textbf{Output:}}] $\mathbf{\boldsymbol{\Phi}}^{*}$ is the value
of $\mathbf{\boldsymbol{\Phi}}$ obtained at the convergence of the
loop.
\end{lyxlist}
\end{algorithm}
\begin{algorithm}[tbh]
\caption{Pseudo code to update $\boldsymbol{\Phi}$ in E-optimal design}

\begin{lyxlist}{00.00.0000}
\item [{\textbf{Input:}}] $\mathbf{H}_{t}$
\end{lyxlist}
\begin{enumerate}
\item Solve the SDP in (\ref{eq:66}) using CVX to update $\mathbf{\boldsymbol{\Phi}}$.
\end{enumerate}
\begin{lyxlist}{00.00.0000}
\item [{\textbf{Output:}}] $\mathbf{\boldsymbol{\Phi}}^{*}$ is the optimal
value obtained using CVX.
\end{lyxlist}
\end{algorithm}

\begin{rem}
The proposed method can be extended for the 3D case, however, the
AOA measurement model for the 3D case cannot be handled in the current
framework as the CRLB expressions in the case of AOA model for the
3D will have sum of two terms (corresponding to the azimuth and elevation
angles) and reformulating the hybrid CRLB matrix into the form similar
to the problem in (\ref{eq:21-2}) is not possible. Nonetheless, we
can extend the proposed algorithm to the reduced hybrid model of TOA-RSS
and in the following we shall discuss the same. The orientation matrix
$\mathbf{J}$ in the 3D case is defined as follows:
\begin{equation}
\mathbf{J}=\begin{bmatrix}\cos\phi_{1}\cos\theta_{1} & \cos\phi_{1}\sin\theta_{1} & \sin\phi_{1}\\
\vdots & \vdots & \vdots\\
\cos\phi_{m}\cos\theta_{m} & \cos\phi_{m}\sin\theta_{1} & \sin\phi_{m}
\end{bmatrix}
\end{equation}
 where $\phi$ is the elevation angle, and the A-optimal design problem
is given by:
\begin{equation}
\begin{aligned}\underset{\mathbf{J}\in\mathcal{D}}{\mathrm{min}}\; & \mathrm{Tr}\left(\left(\mathbf{J}^{T}\mathbf{R}\mathbf{J}\right)^{-1}\right)\end{aligned}
,\label{eq:94}
\end{equation}
where $\mathbf{R}=\mathbf{\Sigma}_{\text{TOA}}^{-1}+\eta^{2}\textbf{D}\mathbf{\Sigma}_{\text{RSS}}^{-1}\textbf{D}.$

The above mentioned problem formulation looks similar to (\ref{eq:21-2})
and the proposed method can be easily adapted to solve for a minimizer
of (\ref{eq:94}).
\end{rem}

\subsection{Computational complexity and proof of convergence}

We first discuss the computational complexity of the proposed algorithm.
The update of the primal variable for the three optimal designs requires
the computation of the matrices $\mathbf{H}_{t}$, $\mathbf{\tilde{\mathbf{A}}}_{1}$,
$\tilde{\mathbf{A}}_{2}$ and $\mathbf{Z}$ (which has a total complexity
of $\mathcal{O}\left(m^{2}n+mn^{2}\right)$) in each iteration. The
update of the dual variable in the A and D optimal designs require
the computations of $\tilde{\boldsymbol{\Phi}}{}_{k}$, $\mathbf{A}$,
$\lambda_{\mathrm{max}}\left(\mathbf{A}\right)$ and $\mathbf{B}$
(which has a total complexity of $\mathcal{O}\left(mn^{2}+n^{3}\right)$),
$\mathbf{C}$ (which has a complexity of $\mathcal{O}\left(m^{2}n\right)$)
and the EVD of $\mathbf{C}$ (which requires $\mathcal{O}\left(n^{3}\right)$
flops) in each inner loop iteration. As $m>n$, the overall complexity
of the proposed algorithm for A and D optimal design is therefore
of the order $\mathcal{O}\left(m^{2}n\right)$. For E-optimal design,
the dual problem is an SDP and the dual variable $\boldsymbol{\Phi}$
is updated using interior point methods. According to \cite{key-51},
the average worst case computation complexity of semi-definite programming
problem is of the order $\mathcal{O}\left(m^{4.5}\right)$. Table
\ref{comp_comp} summarizes the overall computation complexity of
the proposed algorithm using the three different optimal designs.
\begin{table}[th]
\centering \caption{{\footnotesize{}COMPUTATION COMPLEXITY OF THE PROPOSED ALGORITHM FOR
A, D AND E OPTIMAL DESIGNS}}
\label{comp_comp}%
\begin{tabular}{cccc}
\hline 
{\small{}Optimality Criterion} & {\small{}A-optimal} & {\small{}D-optimal} & {\small{}E-optimal}\tabularnewline
\hline 
\hline 
{\small{}Computation Complexity} & {\small{}$\mathcal{O}\left(m^{2}n\right)$} & {\small{}$\mathcal{O}\left(m^{2}n\right)$} & {\small{}$\mathcal{O}\left(m^{4.5}\right)$}\tabularnewline
\hline 
\end{tabular}
\end{table}

We now briefly discuss the proof of convergence of the proposed algorithm.
As the proposed algorithm is a double loop MM algorithm, we will prove
the convergence of the MM updates over the primal variable $\mathbf{H}$
and the dual variable $\mathbf{\Phi}$ separately. Furthermore, since
the convergence of the MM update over $\mathbf{H}$ depends on the
convergence of the MM update over $\mathbf{\Phi}$, we will first
state the following Lemma:
\begin{lem}
\label{lem:9}The iterative steps of the MM update over dual variable
$\mathbf{\Phi}$ in the proposed algorithm converge to the global
minimum of the dual problem (\ref{eq:28}).
\end{lem}
\begin{IEEEproof}
From (\ref{eq:23-1}) we know that the sequence $\left\{ \mathbf{\Phi}_{k}\right\} $
generated by the MM update of dual variable $\mathbf{\Phi}$ will
monotonically decrease the dual objective function $h\left(\mathbf{\Phi}\right)$.
The convergence of the sequence $\left\{ h\left(\mathbf{\Phi}_{k}\right)\right\} $
to a finite value at a stationary point of the dual problem can be
proved using the details in Section II-C of \cite{key-47} and the
references therein. Moreover, since the dual problem is a convex problem,
the stationary point will also be the global minimizer of $h\left(\mathbf{\Phi}\right)$.
\end{IEEEproof}
Now that the convergence of the MM update over dual variable is established
in Lemma \ref{lem:9}, we move on to state the convergence of the
MM update over the primal variable.
\begin{lem}
The iterative steps of the MM update over primal variable $\mathbf{H}$
in the proposed algorithm converge to a KKT point of the primal problem
(\ref{eqn1h}).
\end{lem}
\begin{IEEEproof}
Since, the primal problem is a maximization problem, the sequences
$\left\{ \mathbf{H}_{t}\right\} $ generated by the MM update of primal
variable $\mathbf{H}$ will monotonically increase the objective function
$f\left(\mathbf{H}\right)$. Similar to the details mentioned in Lemma
\ref{lem:9}, the sequence $\left\{ f\left(\mathbf{H}_{t}\right)\right\} $
generated can also be proved to converge to a finite value, and that
will be a KKT point of the primal problem. 
\end{IEEEproof}

\section{NUMERICAL SIMULATIONS AND RESULTS}

In this section, we discuss various numerical simulations under different
noise settings and design parameters to evaluate the performance of
the proposed method. 

\subsection{Convergence of the proposed algorithm}

First, we check the convergence of the proposed algorithm and compare
its performance with the naive way of placing sensors uniformly on
a circle of chosen radius encompassing the target. Let us consider
that there are 4 sensors which are to be optimally placed and the
target is at the centre and the sensors are to be placed on the circle
with radius equal to $10$ meters. For the uniform placement, let
us take the following orientation: 
\begin{equation}
\begin{array}{ll}
\textbf{J}_{\text{uni}}^{T}=\begin{bmatrix}10 & 0 & -10 & 0\\
0 & 10 & 0 & -10
\end{bmatrix}.\end{array}\label{eq:93}
\end{equation}
The noise covariance matrices corresponding to TOA, RSS and AOA are
taken to be non-diagonal with its elements randomly sampled from uniform
distribution $\mathcal{U}\left[0,1\right]$.

In Fig. 2, we show the objective (A, D and E design criteria, for
D-optimal we plot $-\mathrm{log}\left|\left(\mathbf{H}^{T}\mathbf{R}\mathbf{H}\right)\right|$
and for E-optimal we plot $\lambda_{\mathrm{max}}\left(\left(\mathbf{H}^{T}\mathbf{R}\mathbf{H}\right)^{-1}\right)$)
vs iterations\footnote{By iterations we mean the iterations over the primal variable $\mathbf{J}$
or in other words the outer outer loop of our primal dual MM algorithm.}. The primal variable ($\mathbf{J}$) in the proposed algorithm (for
all three designs) is initialized with the uniform configuration in
(\ref{eq:93}) and the variable $\boldsymbol{\Phi}$ (in the case
of A and D-optimal design) is initialized randomly. It can be seen
from the plots in Fig. 2 that the proposed algorithm monotonically
decreases the design objective and eventually converges to an optimum.
The optimal configurations obtained are also shown in Fig. 2. 
\begin{figure*}[tbh]
\centering{}\centering\subfloat[A-optimal design]{\begin{centering}
\begin{minipage}[t]{0.6\columnwidth}%
\begin{center}
\includegraphics[width=3.5cm,height=2.8cm]{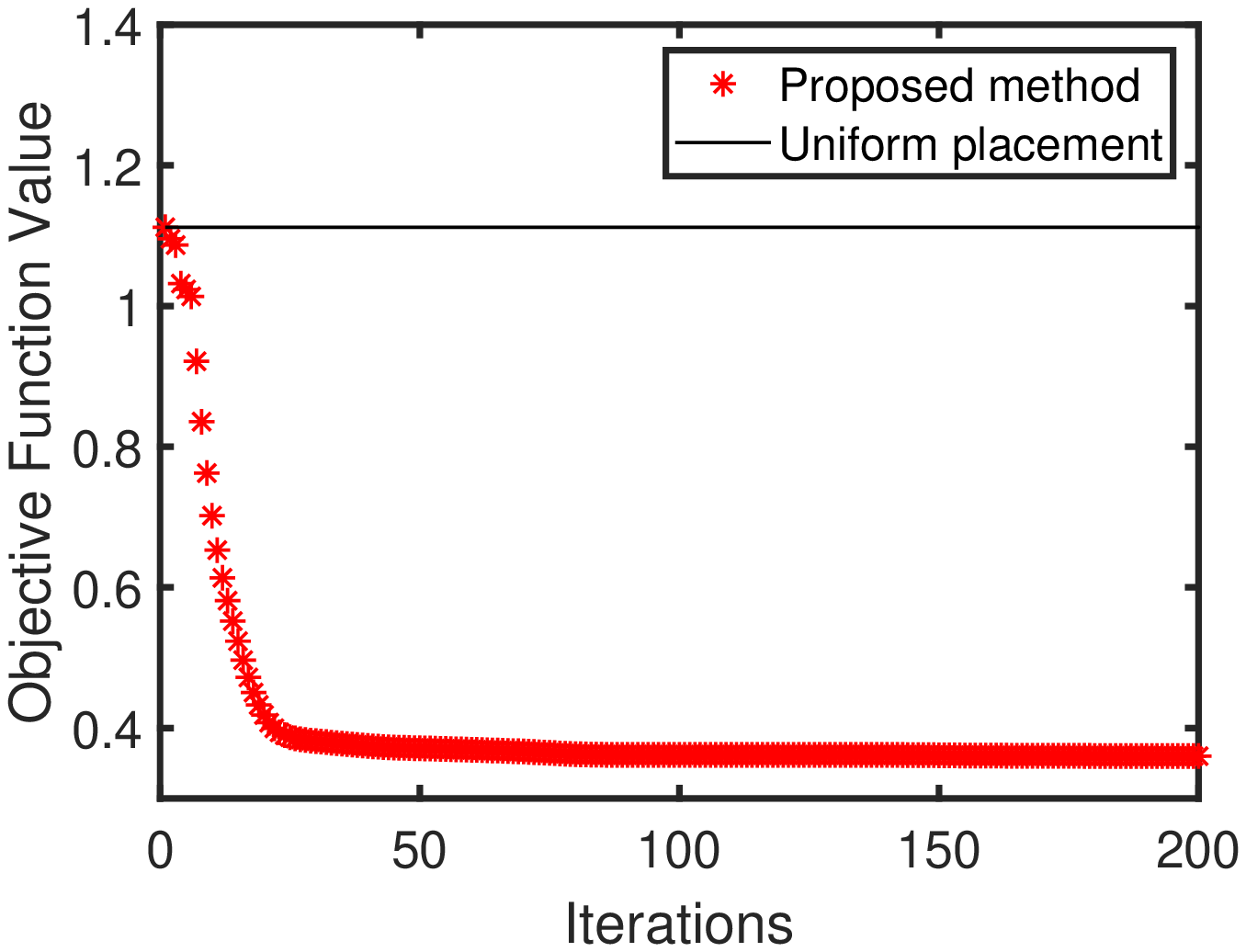}
\par\end{center}
\begin{center}
\includegraphics[width=3.5cm,height=3cm]{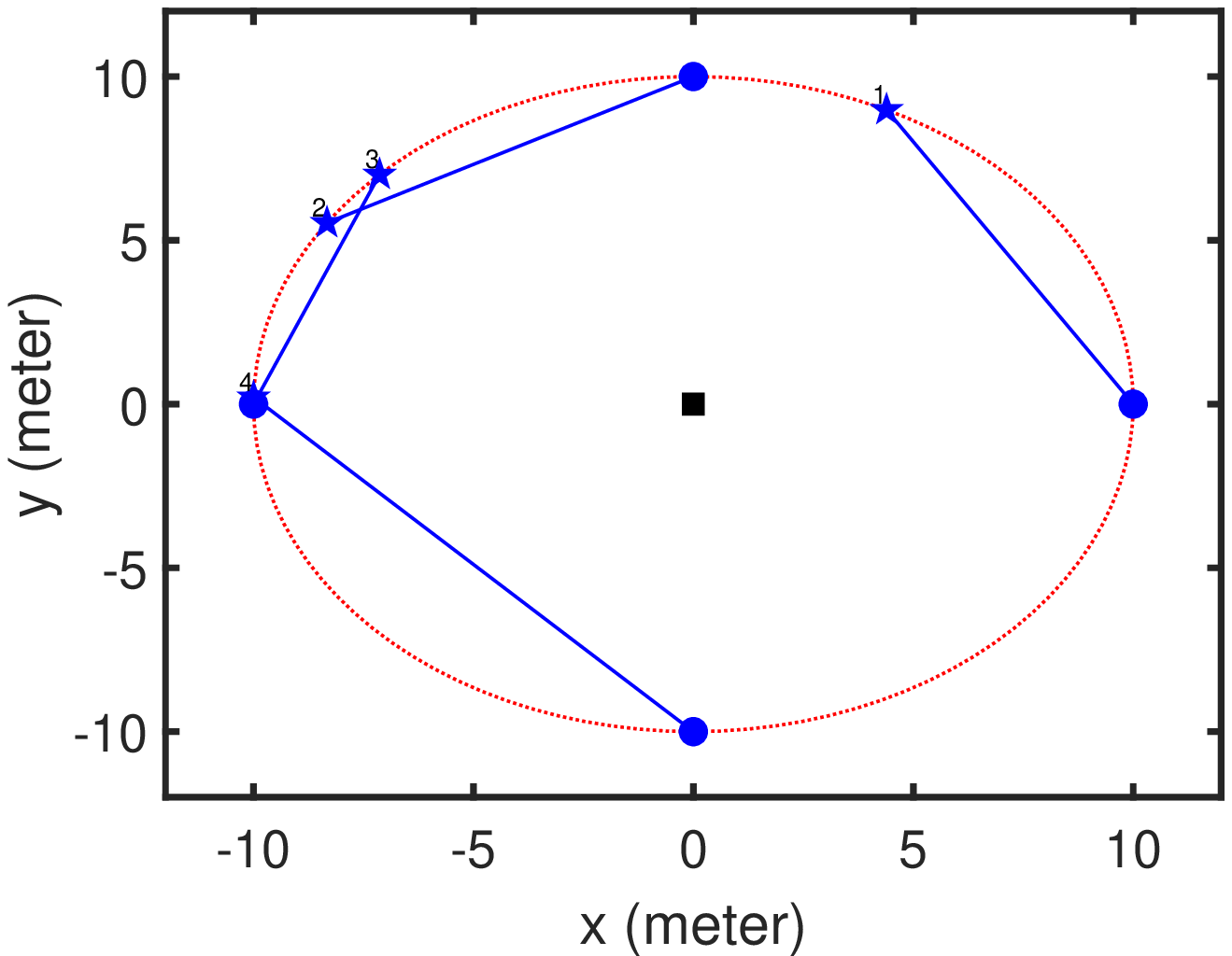}
\par\end{center}%
\end{minipage}
\par\end{centering}
}\subfloat[D-optimal design]{\centering{}%
\begin{minipage}[t]{0.6\columnwidth}%
\begin{center}
\includegraphics[width=3.5cm,height=2.8cm]{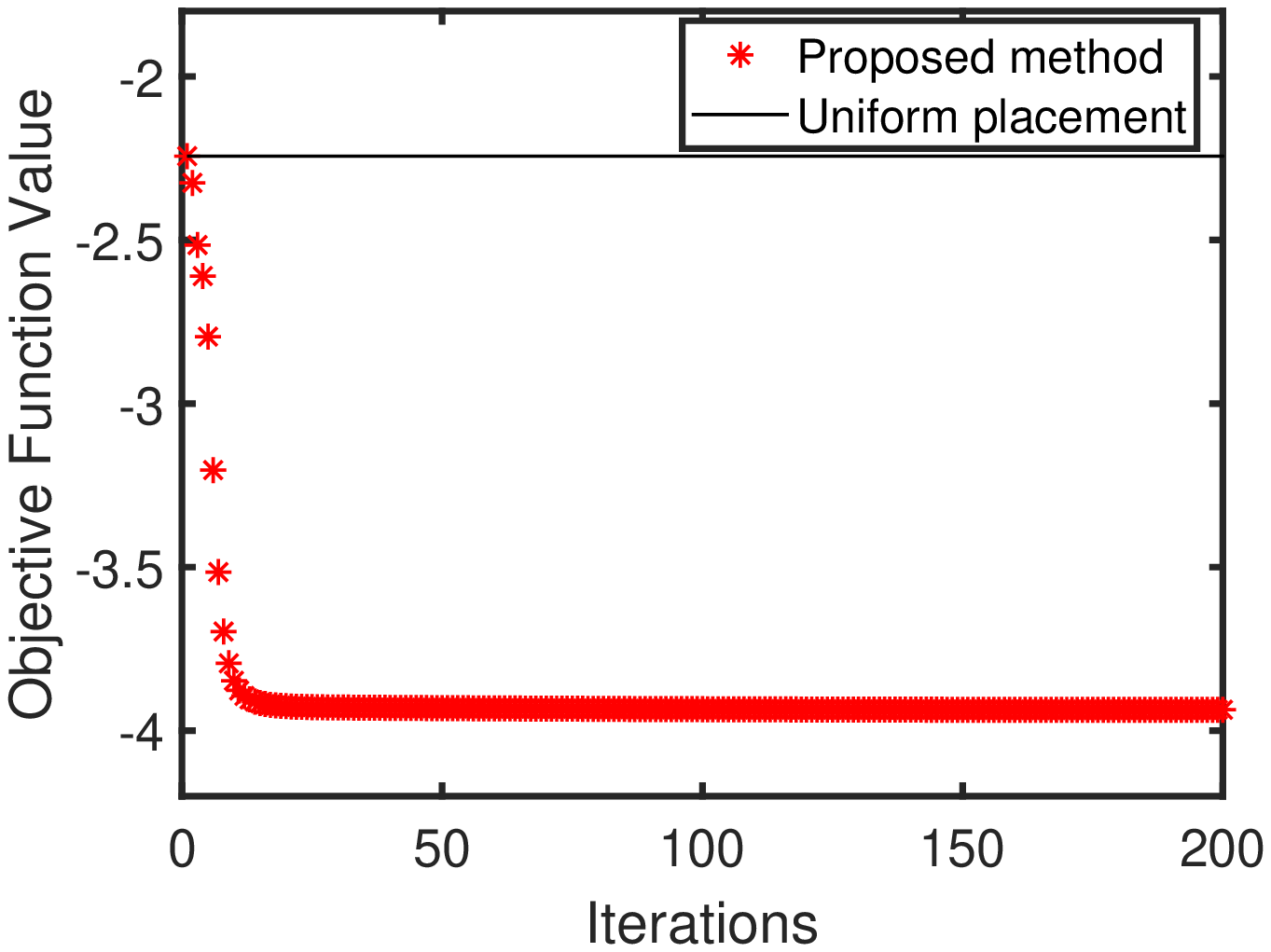}
\par\end{center}
\begin{center}
\includegraphics[width=3.5cm,height=3cm]{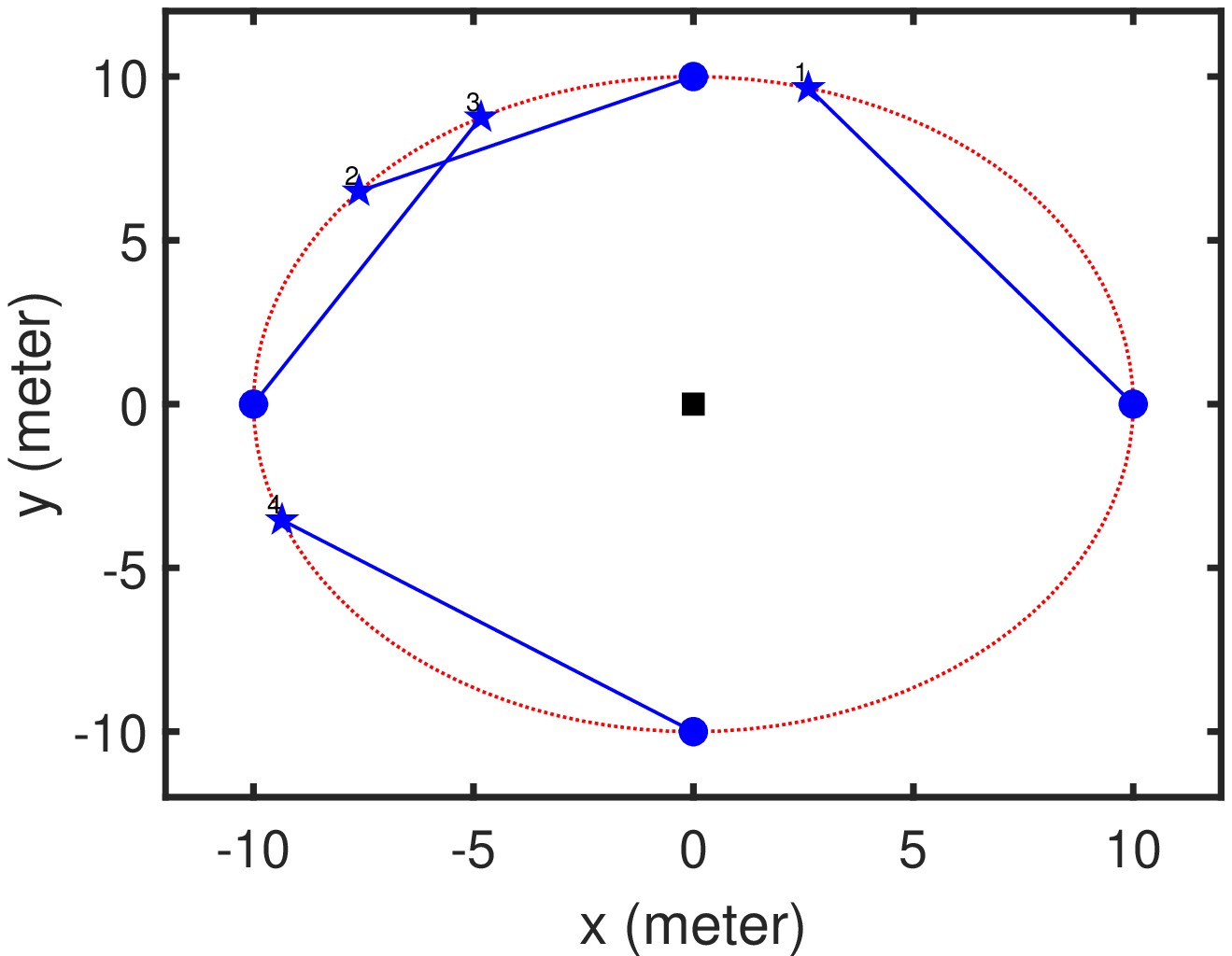}
\par\end{center}%
\end{minipage}}\subfloat[E-optimal design]{\centering{}%
\begin{minipage}[t]{0.6\columnwidth}%
\begin{center}
\includegraphics[width=3.5cm,height=2.8cm]{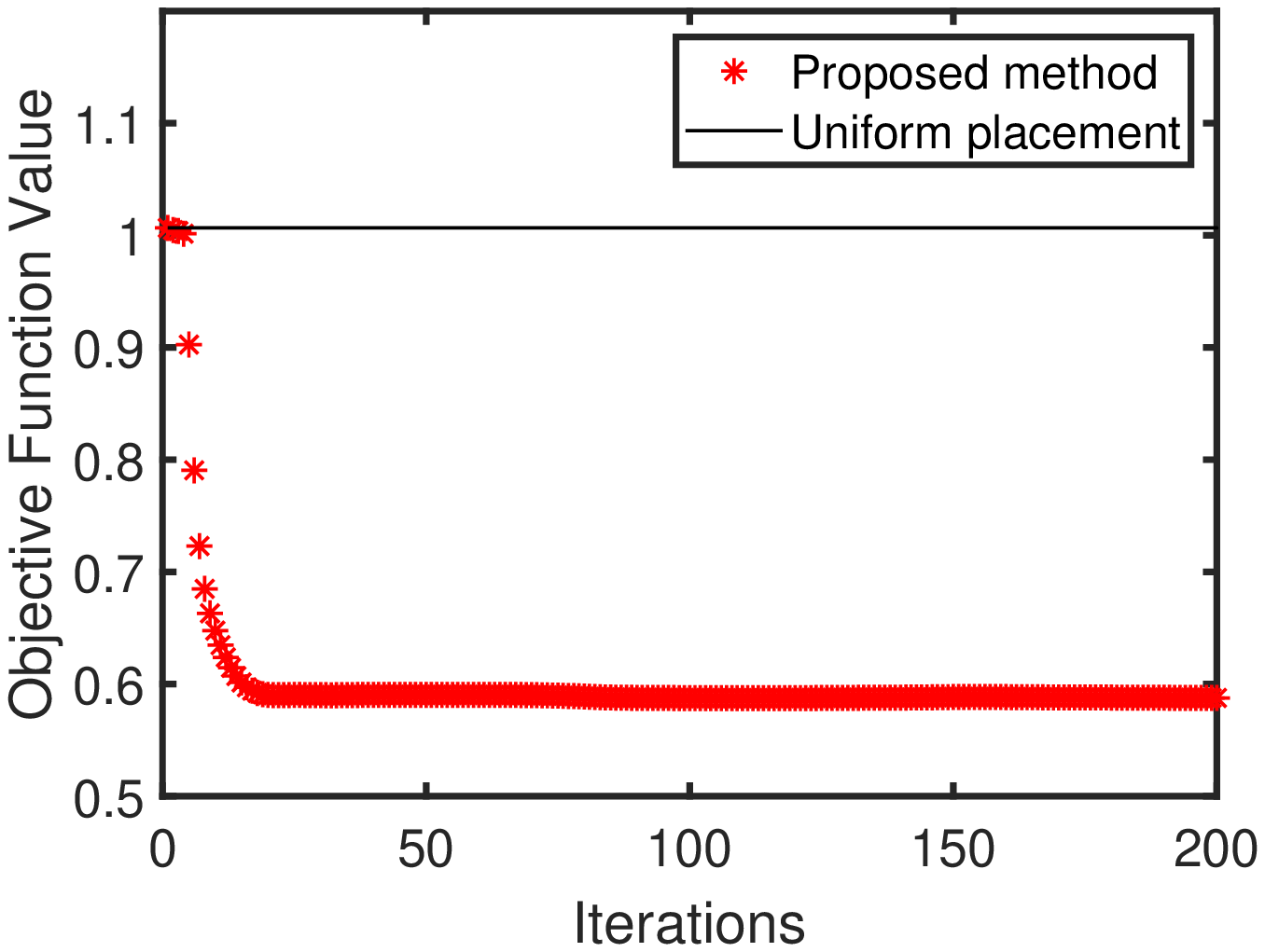}
\par\end{center}
\begin{center}
\includegraphics[width=3.5cm,height=3cm]{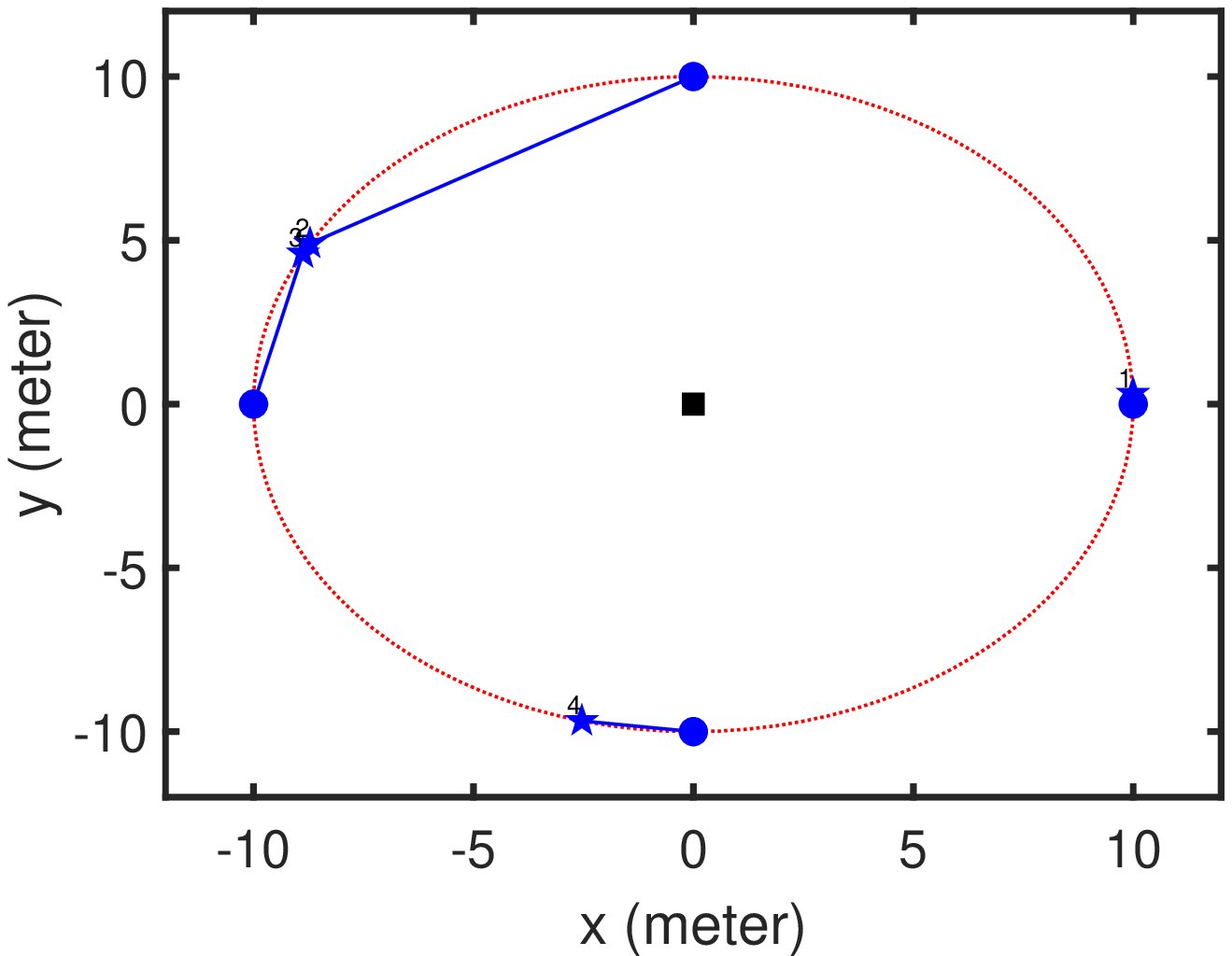}
\par\end{center}%
\end{minipage}}\caption{{\small{}\label{2d view}Convergence plot and corresponding sensor
placement for 2D hybrid TOA-RSS-AOA under correlated noise. Top: convergence
plots; Bottom: sensor placement with black square: target; blue circle:
initial sensors' positions; blue pentagram: final sensors' positions.}}
\end{figure*}

The initialization using the uniform placement also gives a comparison
(in terms of the design objective) between this uniform geometry and
the optimal sensor geometry obtained using the proposed algorithm.
The optimal sensor placement obtained via proposed algorithm using
A, D and E optimal designs show an improvement of $40\%-70\%$ with
respect to the uniform placement. This shows that improvement in localization
accuracy can be achieved using sensor-target geometry obtained by
the proposed algorithm. 

\subsection{Proposed algorithm for the 3D case}

\begin{figure*}[tbh]
\centering{}\centering\subfloat[A-optimal design]{\begin{centering}
\begin{minipage}[t]{0.6\columnwidth}%
\begin{center}
\includegraphics[width=3.5cm,height=2.8cm]{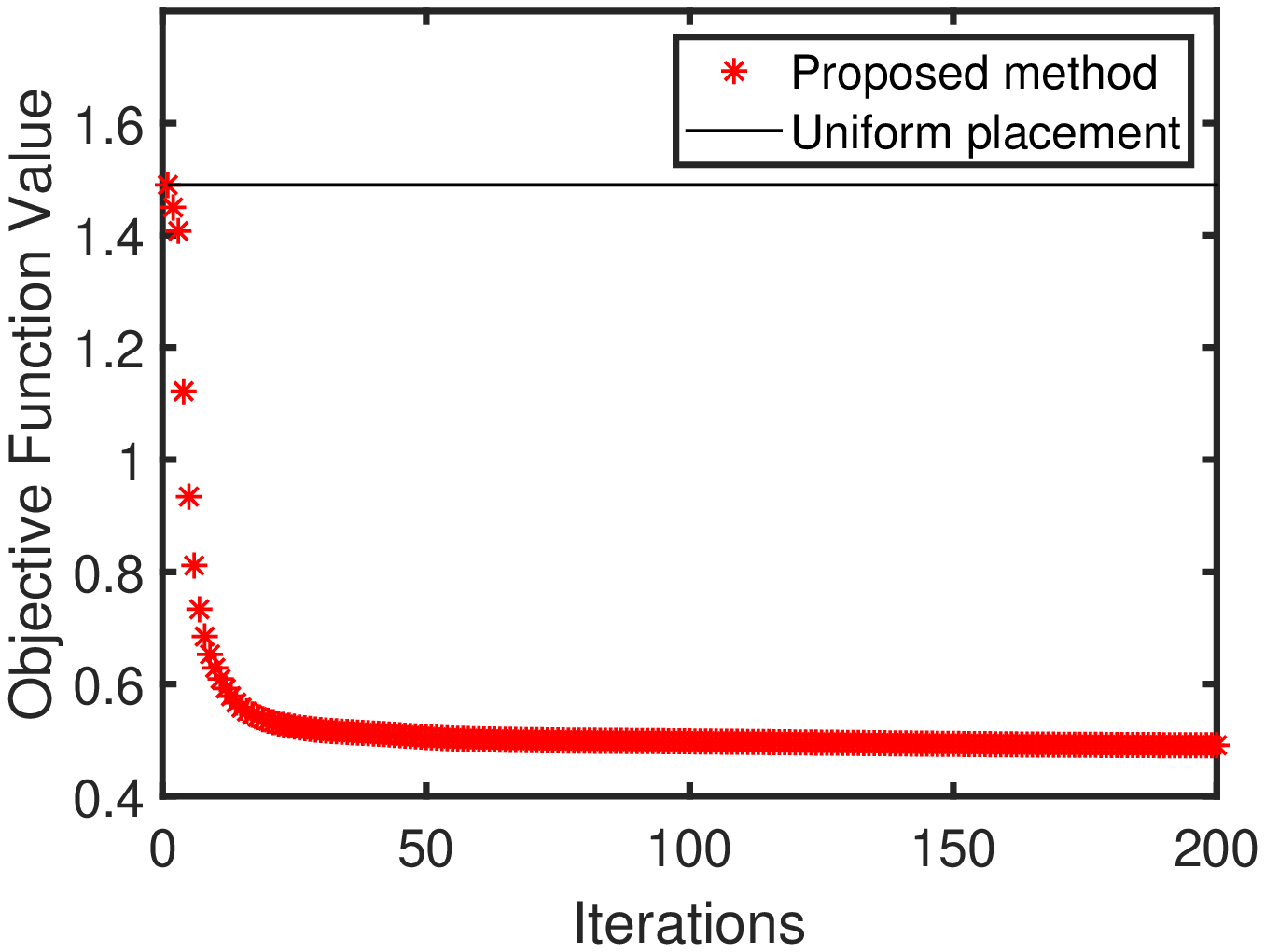}
\par\end{center}
\begin{center}
\includegraphics[width=4cm,height=2.8cm]{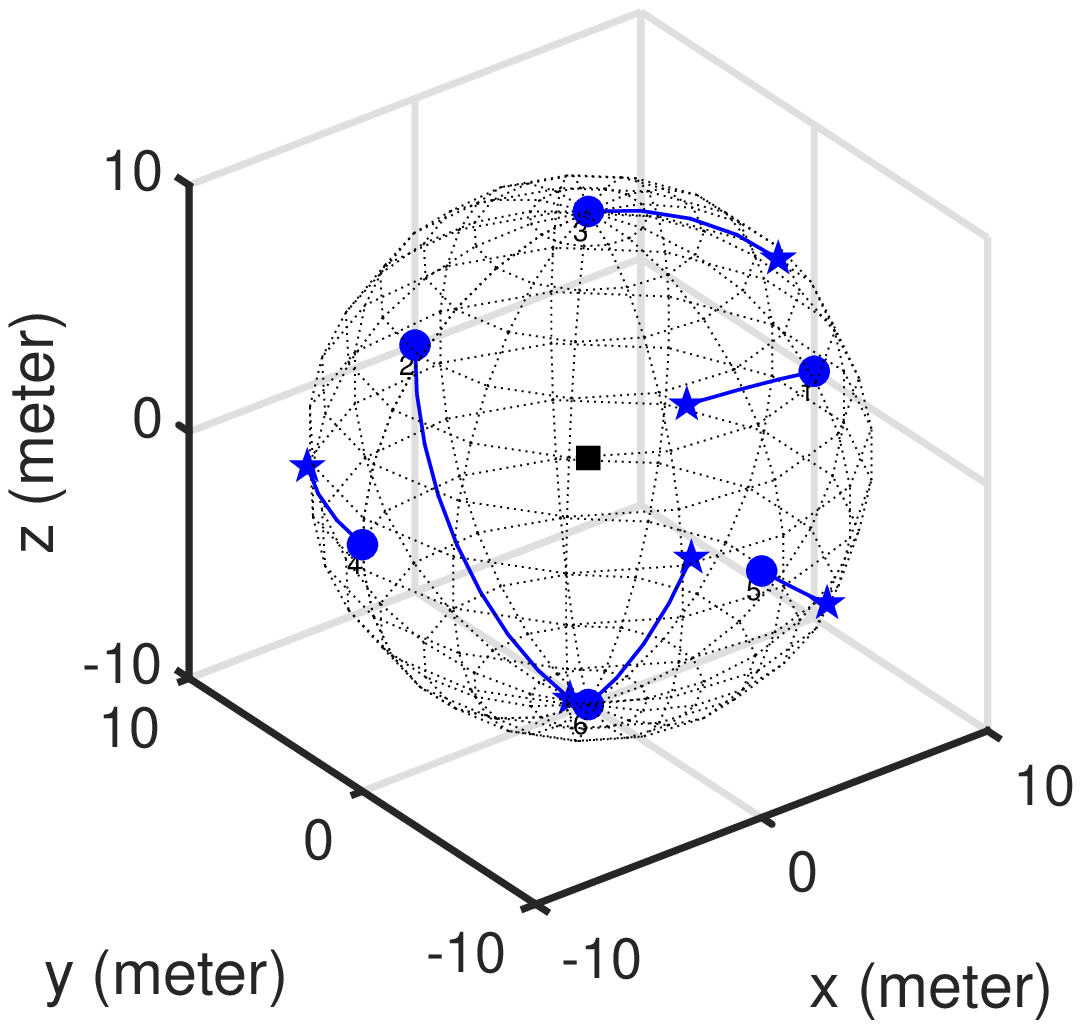} 
\par\end{center}%
\end{minipage}
\par\end{centering}
}\subfloat[D-optimal design]{\centering{}%
\begin{minipage}[t]{0.6\columnwidth}%
\begin{center}
\includegraphics[width=3.5cm,height=2.8cm]{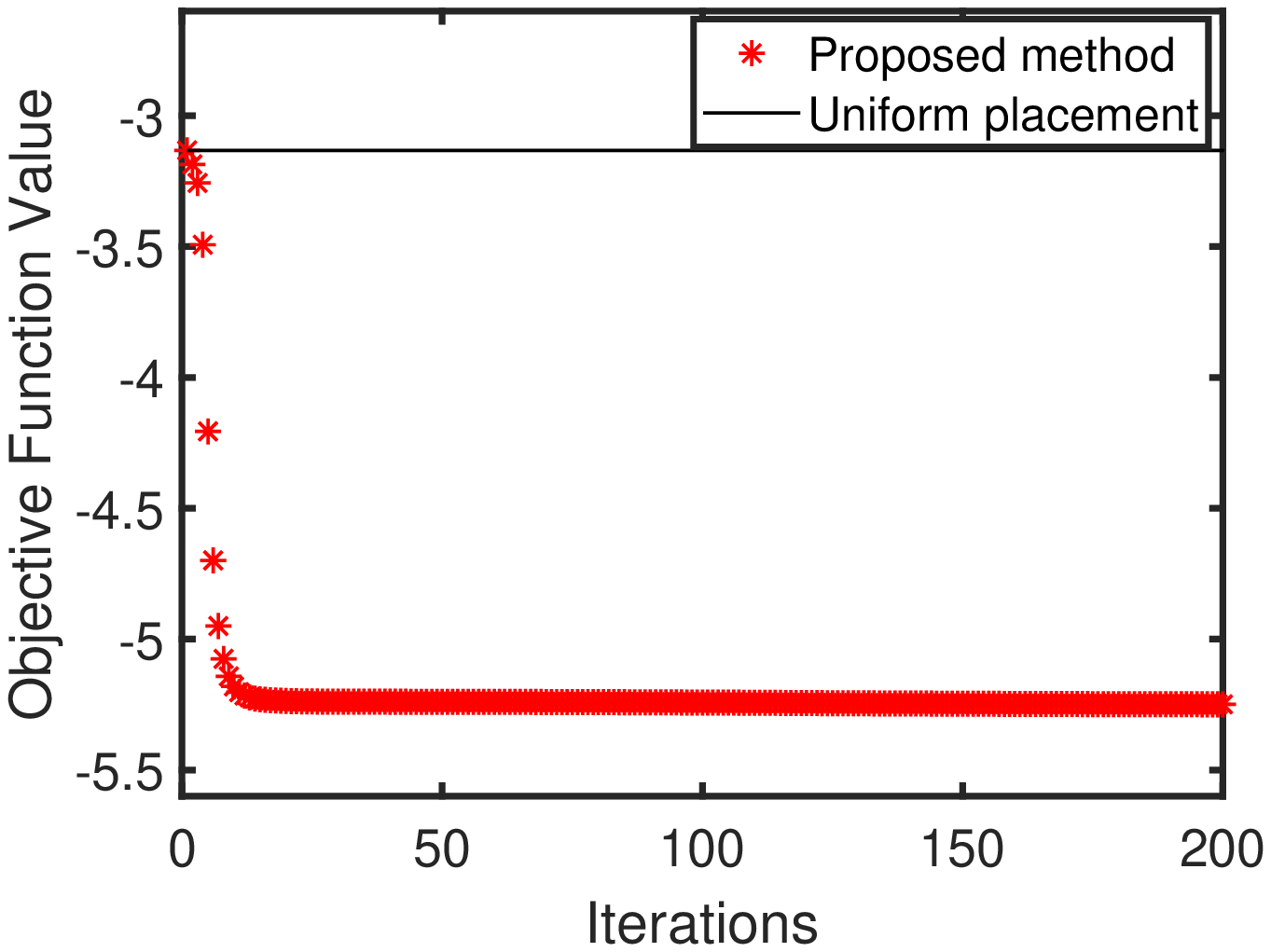}
\par\end{center}
\begin{center}
\includegraphics[width=4cm,height=2.8cm]{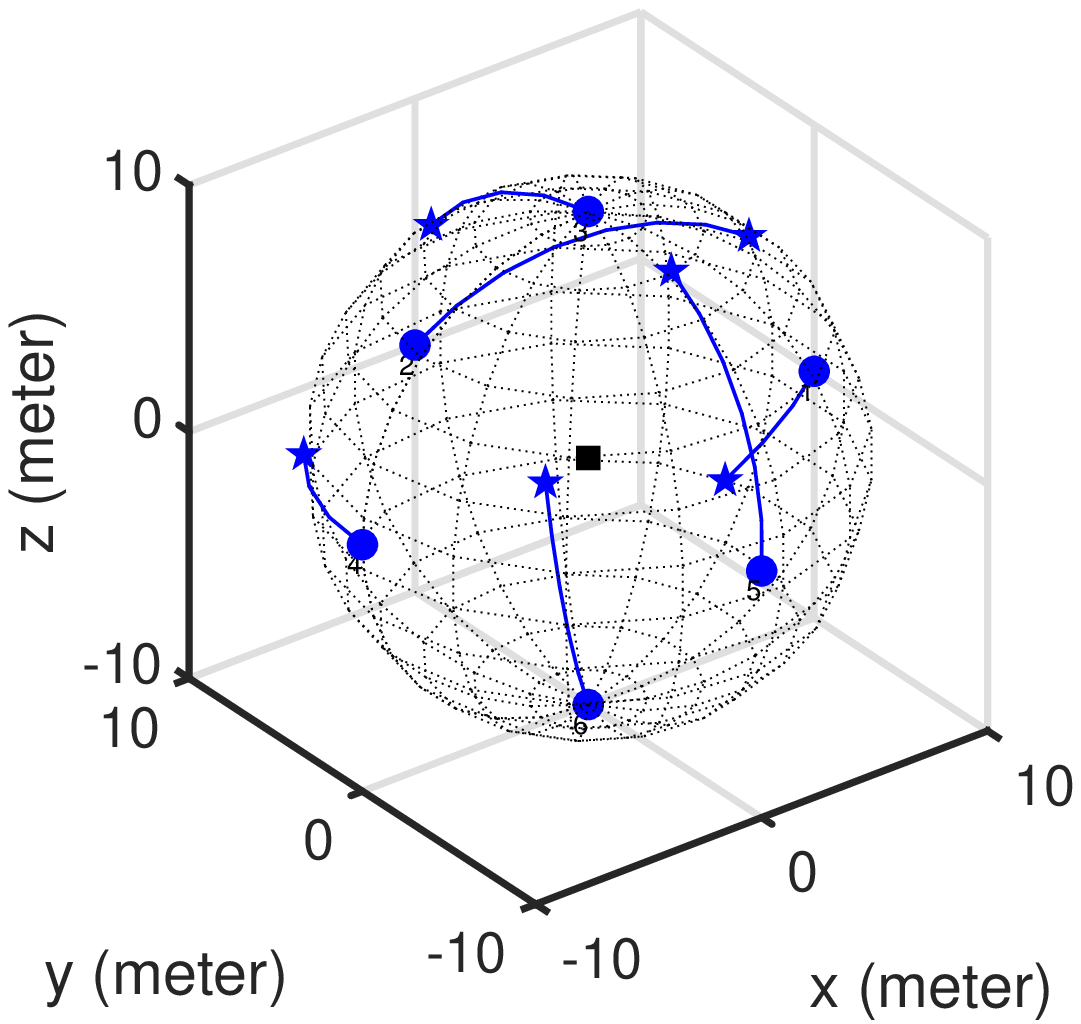}
\par\end{center}%
\end{minipage}}\subfloat[E-optimal design]{\centering{}%
\begin{minipage}[t]{0.6\columnwidth}%
\begin{center}
\includegraphics[width=3.5cm,height=2.8cm]{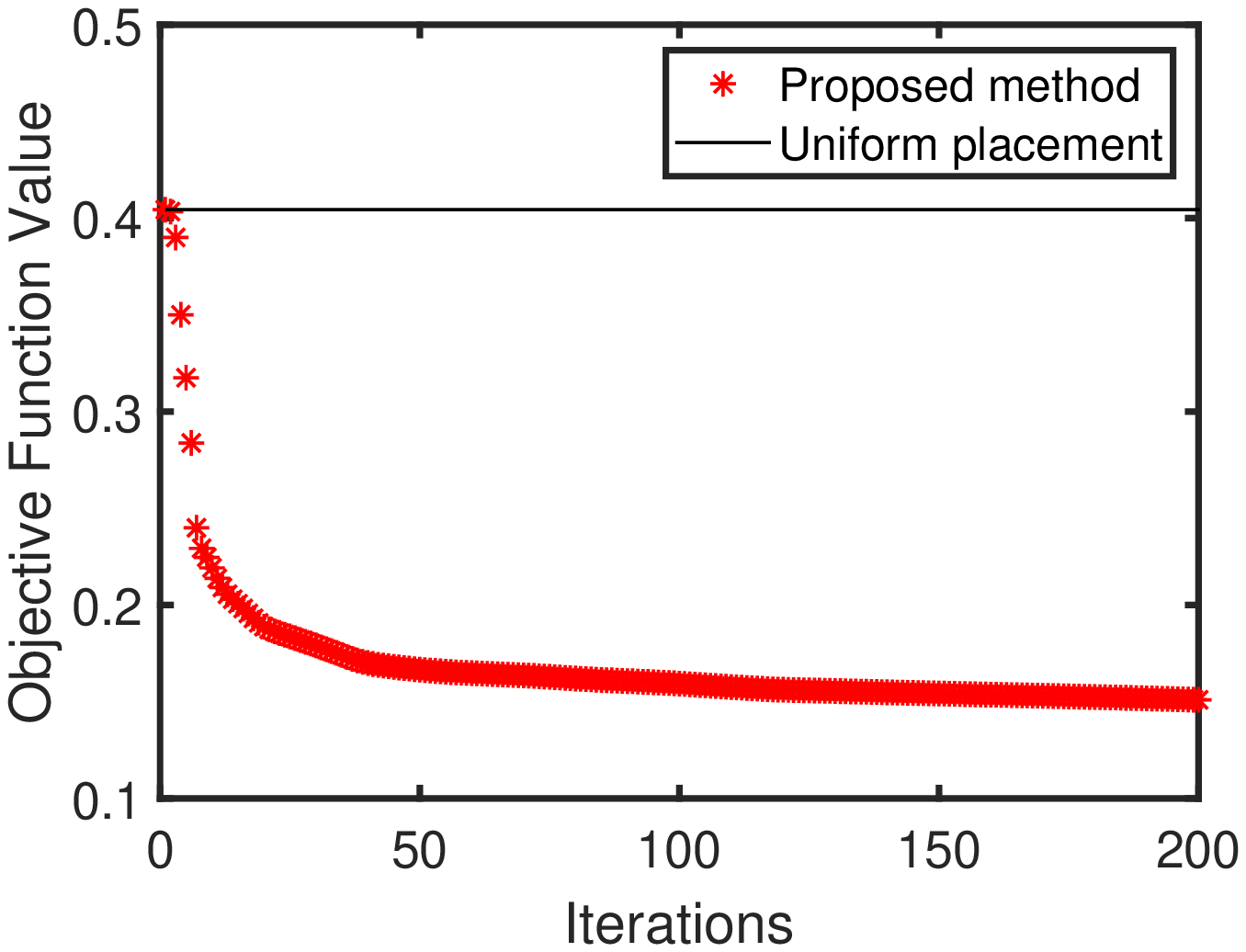}
\par\end{center}
\begin{center}
\includegraphics[width=4cm,height=2.8cm]{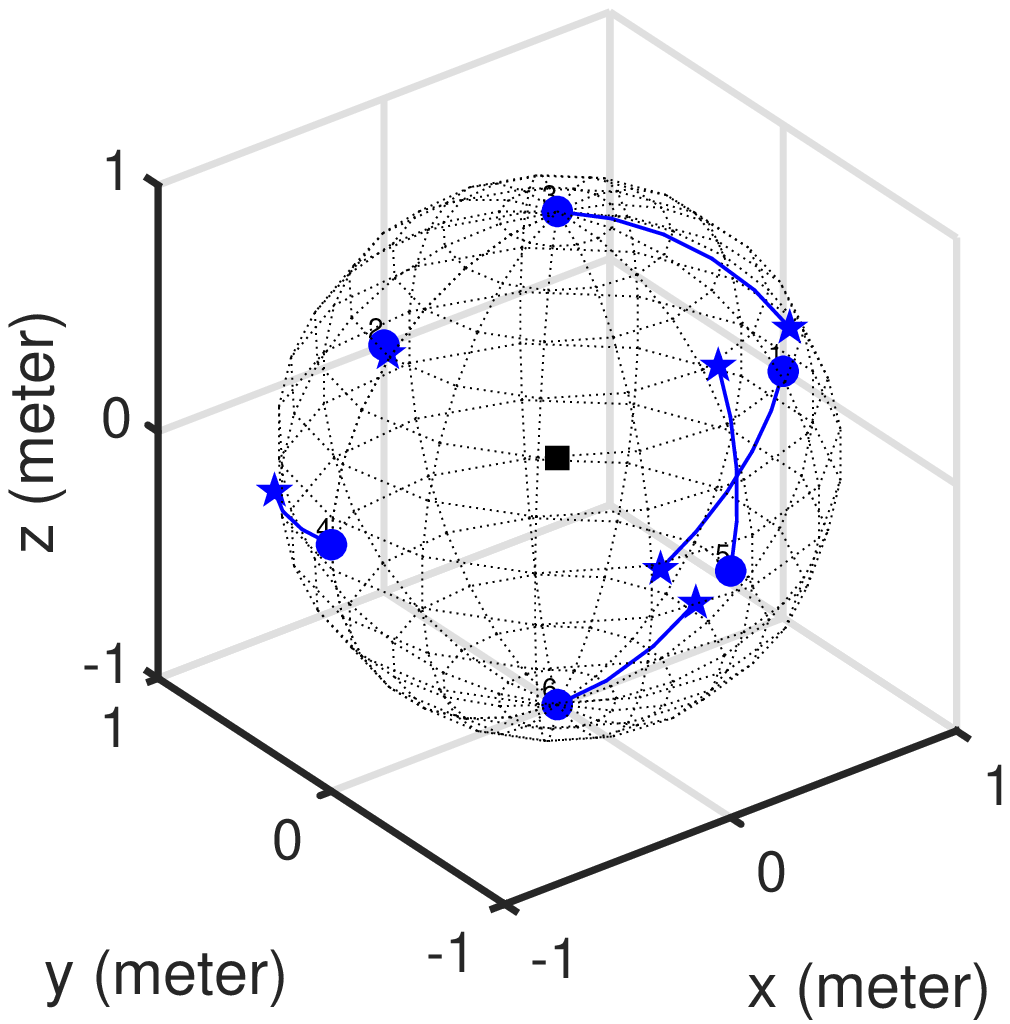}
\par\end{center}%
\end{minipage}}\caption{{\small{}\label{3d view}Convergence plot and corresponding sensor
placement for 3D hybrid TOA-RSS under correlated noise. Top: convergence
plots; Bottom: sensor placement with black square: target; blue circle:
initial sensors' positions; blue pentagram: final sensors' positions.}}
\end{figure*}

For the 3D case, we have undertaken the design of placing 6 sensors
on the surface of sphere of radius $10$ meters with target at its
center. The initialization used for our proposed method is given by:
\begin{equation}
\begin{array}{ll}
\textbf{J}_{\text{uni}}^{T}=\begin{bmatrix}10 & 0 & 0 & -10 & 0 & 0\\
0 & 10 & 0 & 0 & -10 & 0\\
0 & 0 & 10 & 0 & 0 & -10
\end{bmatrix}.\end{array}
\end{equation}
As remarked in the algorithmic section, the proposed method here is
only for hybrid TOA-RSS model. The measurements are corrupted with
correlated noise of zero-mean and positive definite covariance matrices
which are generated similar to the 2D case. 

The proposed method is initialized using the uniform placement and
the optimal sensor placements are obtained. The variation of objective
with iterations are shown in Fig. \ref{3d view} for all three designs.
These plots show monotonic decrease in the objectives which validates
the convergence of the proposed algorithm for the 3D case. 

\subsection{Performance in the presence of uncorrelated noise measurements}

Next, we will study the performance of the proposed algorithm for
the uncorrelated noise case\footnote{From hereafter, the results shown are only for A-optimal criterion.
We could not include the results for D and E-optimal criteria due
to lack of space.}. To this end, we have assumed that the sensors are to be placed such
that the distance from the target (at origin) is always one meter
and the measurements are affected by diagonal noise covariance matrices
given by: 
\begin{equation}
\begin{array}{ll}
\mathbf{\Sigma}_{\text{TOA}}=\text{diag}[\gamma^{2},\dots,\gamma^{2}]\\
\mathbf{\Sigma}_{\text{RSS}}=\text{diag}[\sigma^{2},\dots,\sigma^{2}]\\
\mathbf{\Sigma}_{\text{AOA}}=\text{diag}[\tau^{2},\dots,\tau^{2}]
\end{array}
\end{equation}
For the uniform noise case, the optimal orientation matrix satisfies
the following relation \cite{key-40}: 
\begin{equation}
\textbf{J}_{*}^{T}\textbf{J}_{*}=\frac{m}{2}\textbf{I}_{n}.\label{eq:76}
\end{equation}

\begin{table}[!th]
\centering \caption{{\footnotesize{}COMPARISON OF THEORETICAL AND NUMERICAL VALUE OF TRACE
OF CRLB}}
\label{tvalue} %
\begin{tabular}{ccc}
\hline 
{\small{}No of sensors} & {\small{}$\mathrm{Tr}\left(\mathbf{C}_{\mathrm{theo}}\right)$} & {\small{}$\mathrm{Tr}\left(\mathbf{C}_{\mathrm{algo}}\right)$}\tabularnewline
\hline 
\hline 
{\small{}2} & {\small{}0.0959} & {\small{}0.0959}\tabularnewline
{\small{}5} & {\small{}0.0383} & {\small{}0.0383}\tabularnewline
{\small{}10} & {\small{}0.0192} & {\small{}0.0192}\tabularnewline
{\small{}15} & {\small{}0.0128} & {\small{}0.0128}\tabularnewline
\hline 
\end{tabular}
\end{table}
The corresponding theoretical value of CRLB can also be easily calculated
for this special case using a generalized linear inequality as shown
in \cite{key-43}. This theoretical value is given by: 
\begin{equation}
\mathrm{Tr}\left(\mathbf{C}_{\text{theo}}\right)=\frac{4}{\sum\limits _{i=1}^{m}(\frac{\eta^{2}}{d_{i}^{2}\sigma^{2}}+\frac{1}{d_{i}^{2}\tau^{2}}+\frac{1}{\gamma^{2}})}\label{crlb_th}
\end{equation}
In the following, we present the value of the trace of CRLB matrix
obtained at convergence of our algorithm for different values of `$m$'
with the choice of uniform noise variances ($\gamma^{2}=1,\;\sigma^{2}=1,\;\tau^{2}=1$).
All the sensor-target ranges are taken to be equal ($d_{i}=1\;\forall i$).
The trace of CRLB matrix obtained by the proposed algorithm ($\mathrm{Tr}\left(\mathbf{C}_{\text{algo}}\right)$)
and the trace of theoretical CRLB matrix ($\mathbf{\mathrm{Tr}\left(\mathbf{C}_{\text{theo}}\right)}$)
using \eqref{crlb_th} for different number of sensors are listed
in table \ref{tvalue}. From the table, it can be seen that the CRLB
value obtained using the proposed algorithm is equivalent to the theoretical
CRLB value proving the optimality of the proposed algorithm. The optimal
orientation matrix obtained via our algorithm also satisfied the relation
in (\ref{eq:76}).

\subsubsection{\label{subsec:4.3.1}Optimal design for uniform sensor-target ranges}

\begin{figure}[tbh]
\centering\subfloat[]{\begin{centering}
\includegraphics[width=4cm,height=3.5cm]{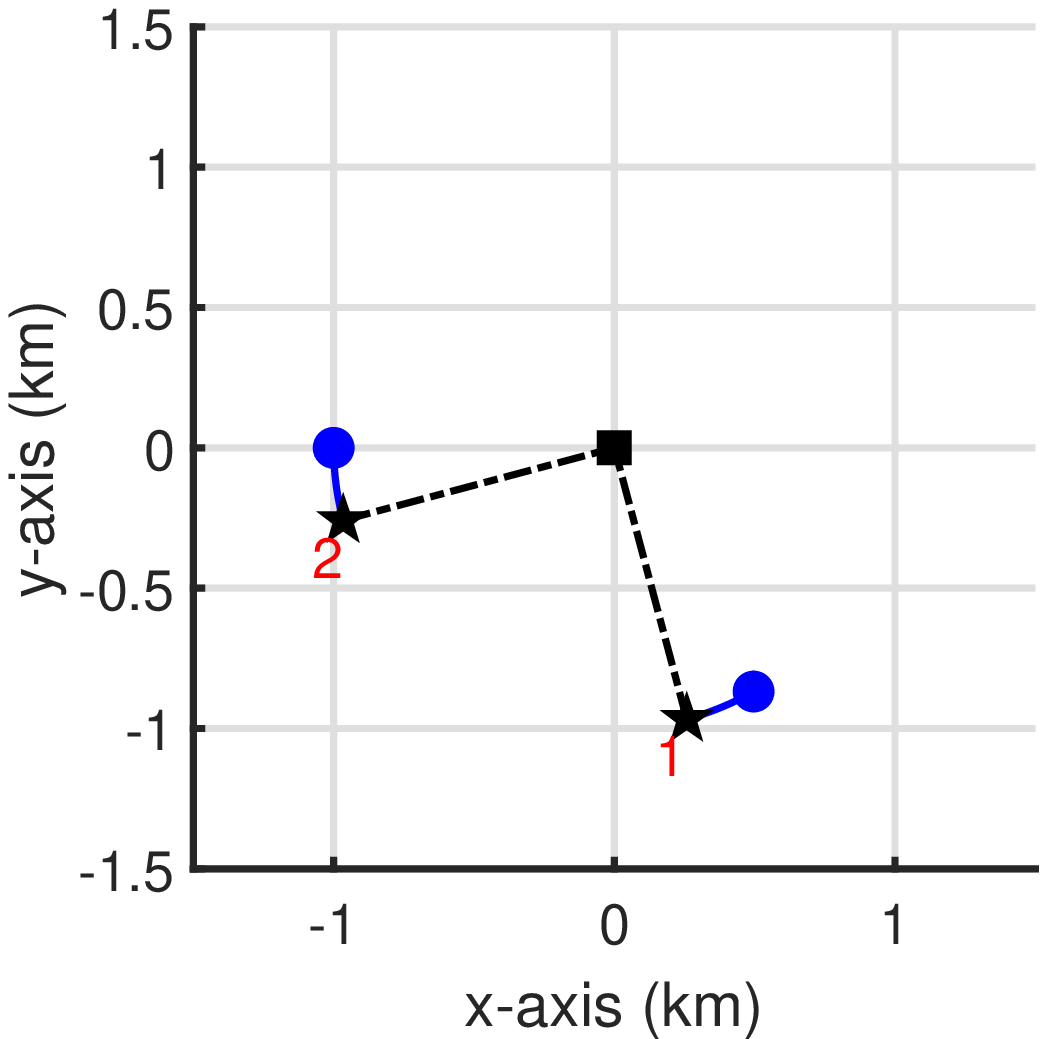} 
\par\end{centering}
}\subfloat[]{\begin{centering}
\includegraphics[width=4cm,height=3.5cm]{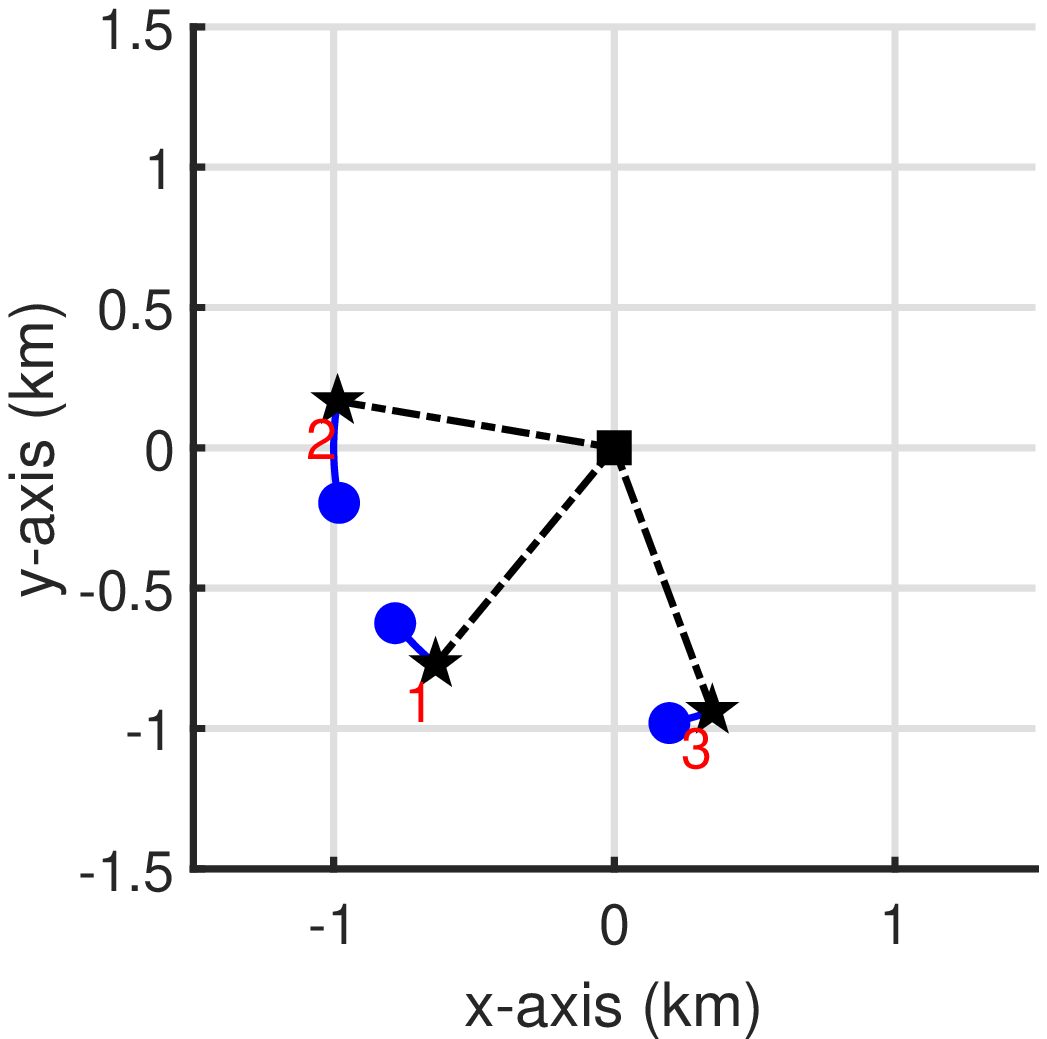} 
\par\end{centering}
}

\caption{{\small{}\label{dia2}Optimal sensor trajectories of (a) example 1
and (b) example 2, $\circ:\text{ sensor initial position ,}*:\text{ sensor final position ,}\square:\text{ target position }$}}
\end{figure}

In this subsection, we assume the distances between the sensors and
the target are equal ($d_{i}$$=d$) and generate the optimal orientations
for different values of $d$. We have considered two different configuration
examples. In example 1, we assume $m=2,\;d=1000\;\mathrm{meter},\;\gamma_{i}=\gamma=1\;\mathrm{meter},\;\sigma_{i}=\sigma=1W,\;\tau_{i}=\tau=1\text{ rad},\;\eta=-4.343$
and the initial positions (used to initialize our algorithm) for the
two sensors are taken as $(0.4,-0.7)\text{ kilometers and }(-1,0)$
kilometers. In example 2, we have assumed 3 sensors with initial positions
given by $(-0.75,-0.6)\text{ kilometers and }(-1,-0.2)\text{ kilometers}$
and $(0.2,-1)$ kilometers. The initial and final sensors positions
(obtained via our proposed algorithm) for example 1 and 2 are shown
in Fig. \ref{dia2}. The optimal value of the A optimal objective
obtained is $2$ sq. meters and $1.333$ sq. meters which is also
the theoretical minimum CRLB value calculated using \eqref{crlb_th}. 

\begin{figure}[tbh]
\begin{centering}
\includegraphics[width=4.2cm,height=3.5cm]{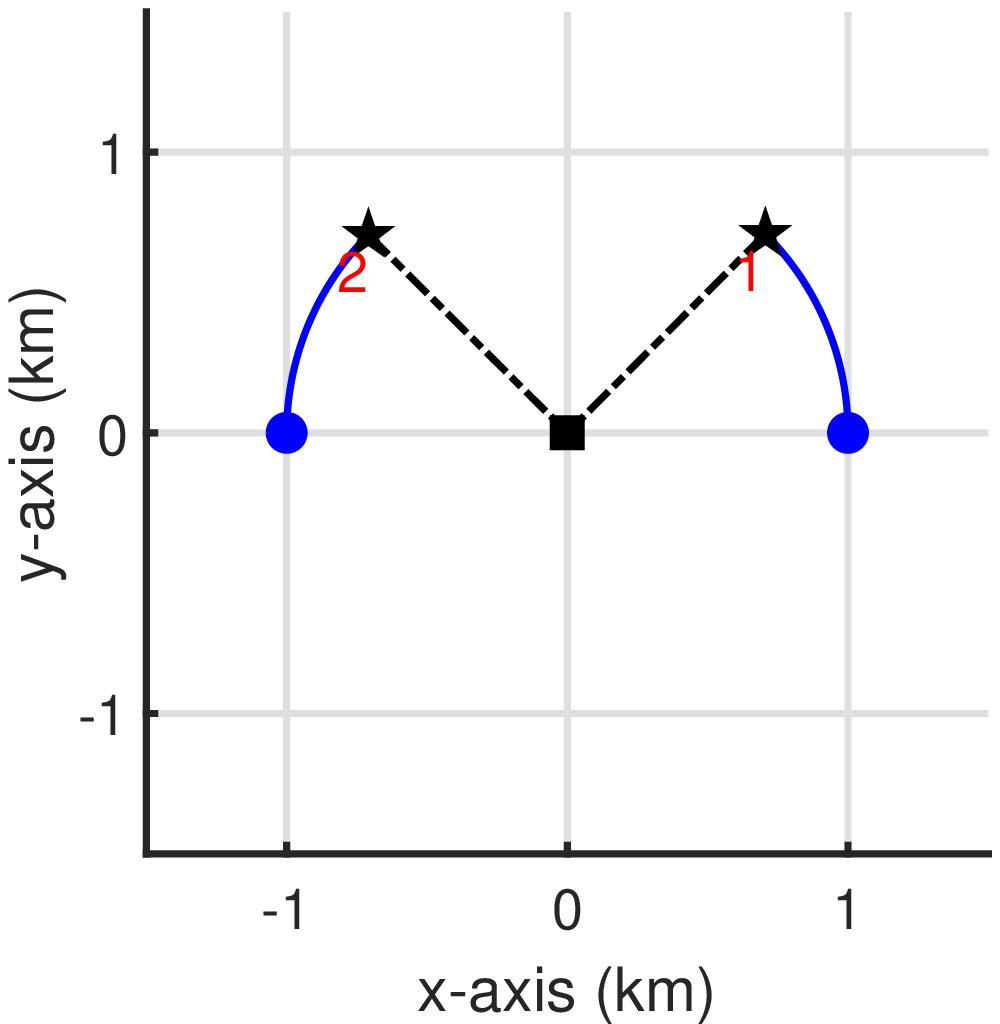}\includegraphics[width=4.2cm,height=3.5cm]{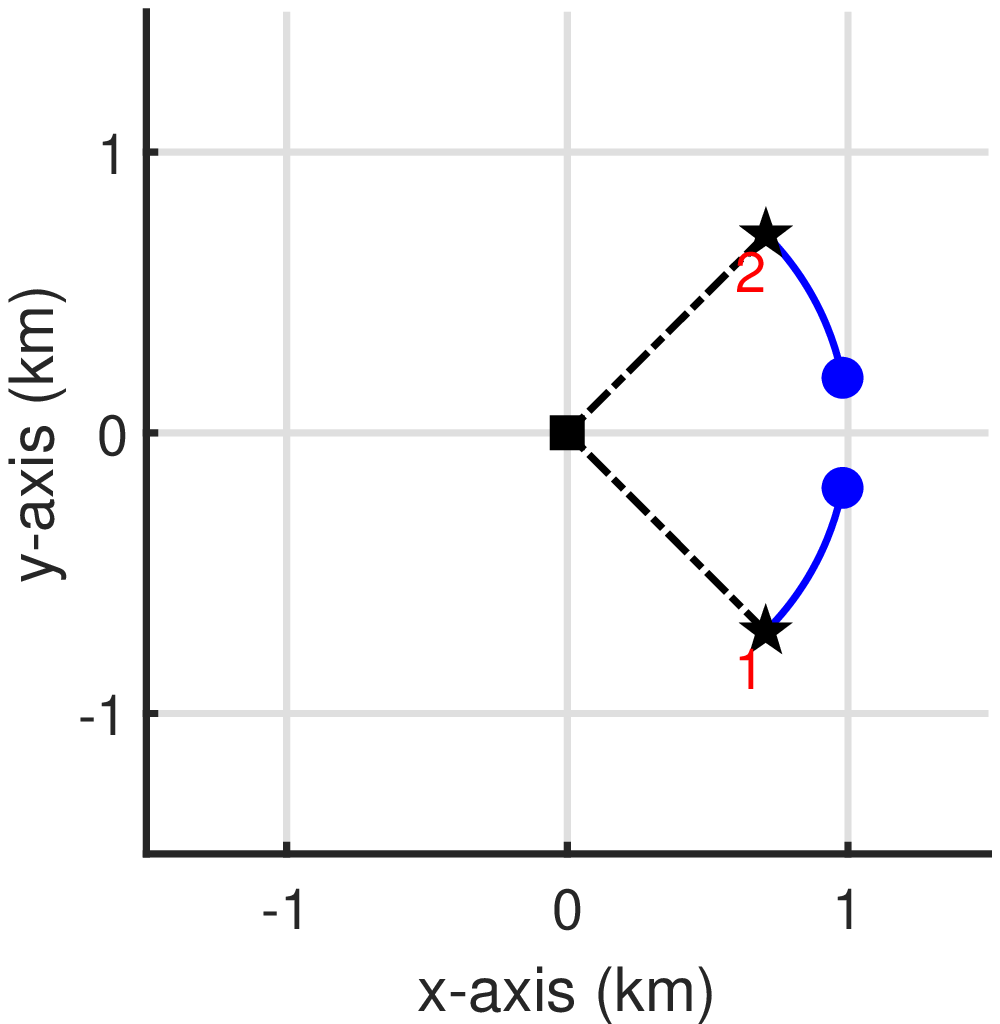}
\par\end{centering}

\caption{\label{dia3}Optimal sensor trajectories and the final geometries
with different sensor initial positions for $m=2$ and uniform $d_{i}$
considered in subsection \ref{subsec:4.3.1} .}
\end{figure}

\begin{figure}[tbh]
\begin{centering}
\includegraphics[width=4.2cm,height=3.5cm]{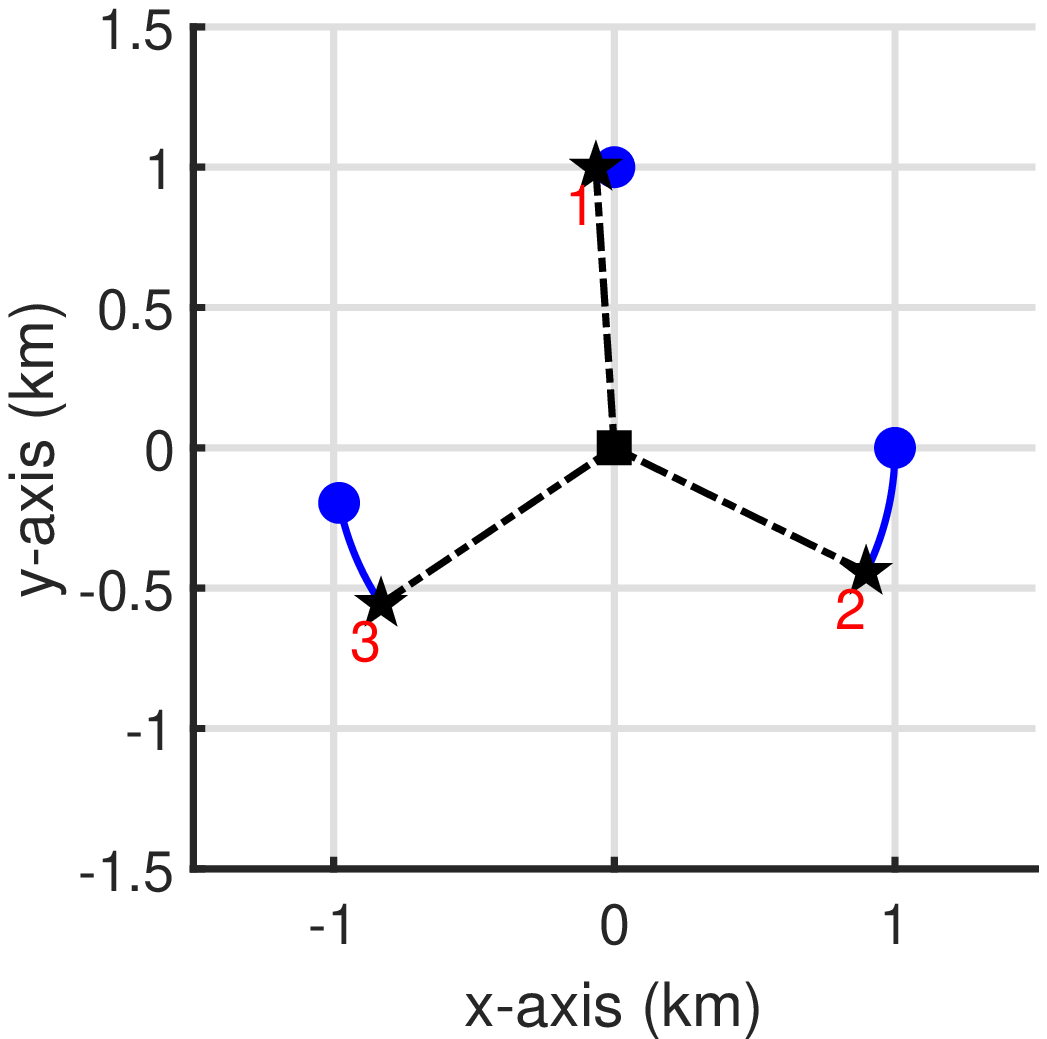}\includegraphics[width=4.2cm,height=3.5cm]{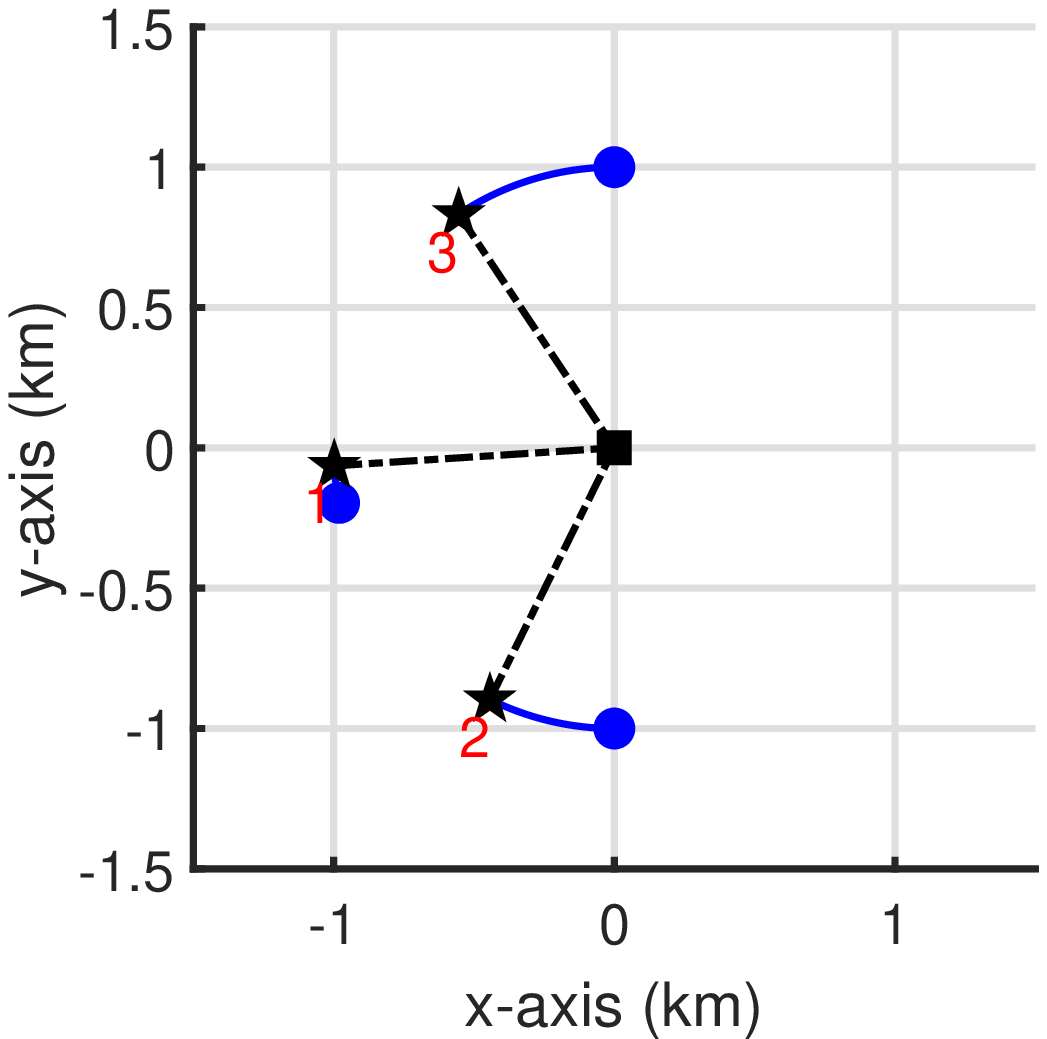}
\par\end{centering}
\caption{\label{dia4}{\small{}Optimal sensor trajectories and the final geometries
with different sensor start positions for $m=3$ and uniform $d_{i}$
}considered in subsection \ref{subsec:4.3.1}{\small{}.}}

\end{figure}

Next, we have designed the optimal geometries for different initial
positions of the sensors for the parameter settings mentioned in example
1 and 2. The optimal orientations obtained via our proposed approach
are shown in Fig. \ref{dia3} and \ref{dia4}. Some of the optimal
configurations obtained via our proposed approach (in Fig \ref{dia3}
and \ref{dia4}) are same as the configurations obtained via the analytical
approach of \cite{key-43}.

\subsubsection{\label{subsec:4.3.2}Optimal design for nonuniform sensor-target
range}

\begin{figure}[tbh]
\begin{centering}
\includegraphics[width=4.2cm,height=3.5cm]{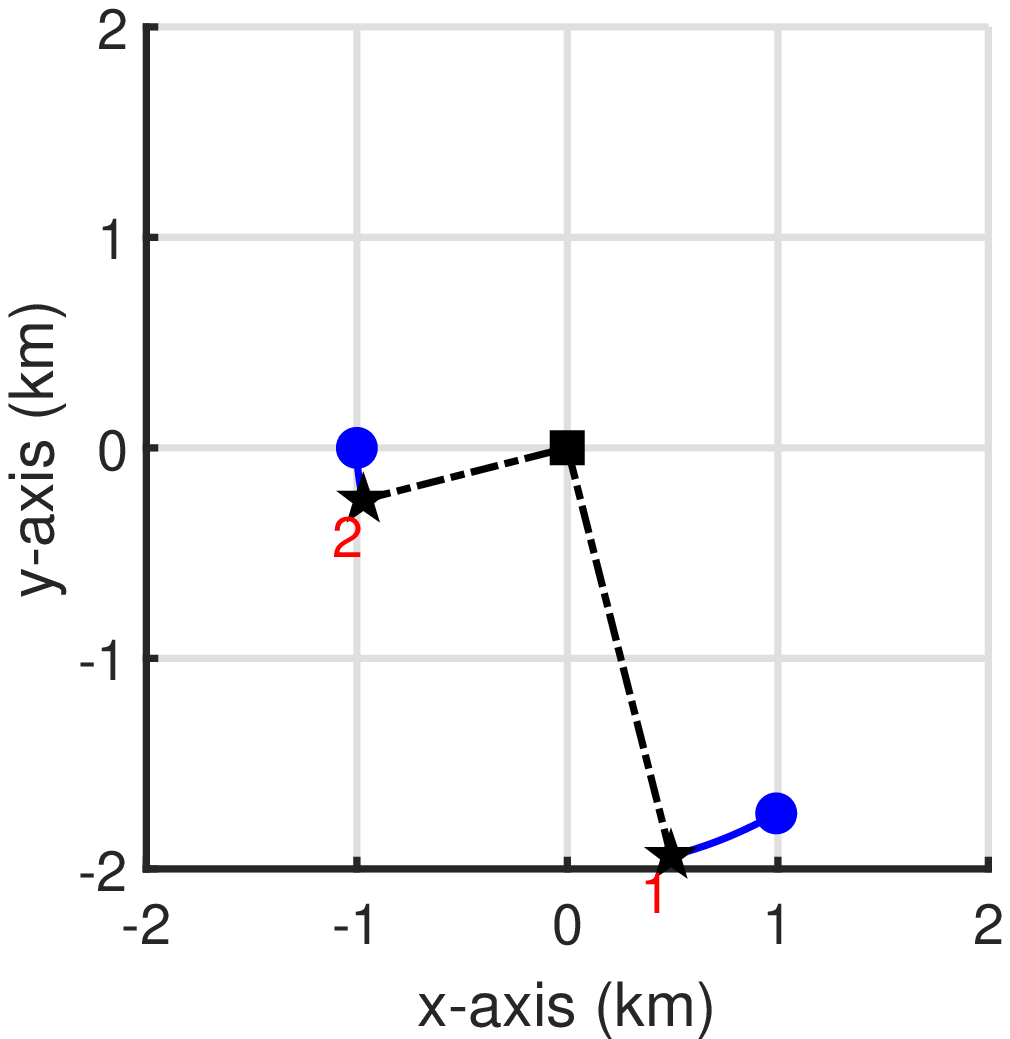}\includegraphics[width=4.2cm,height=3.5cm]{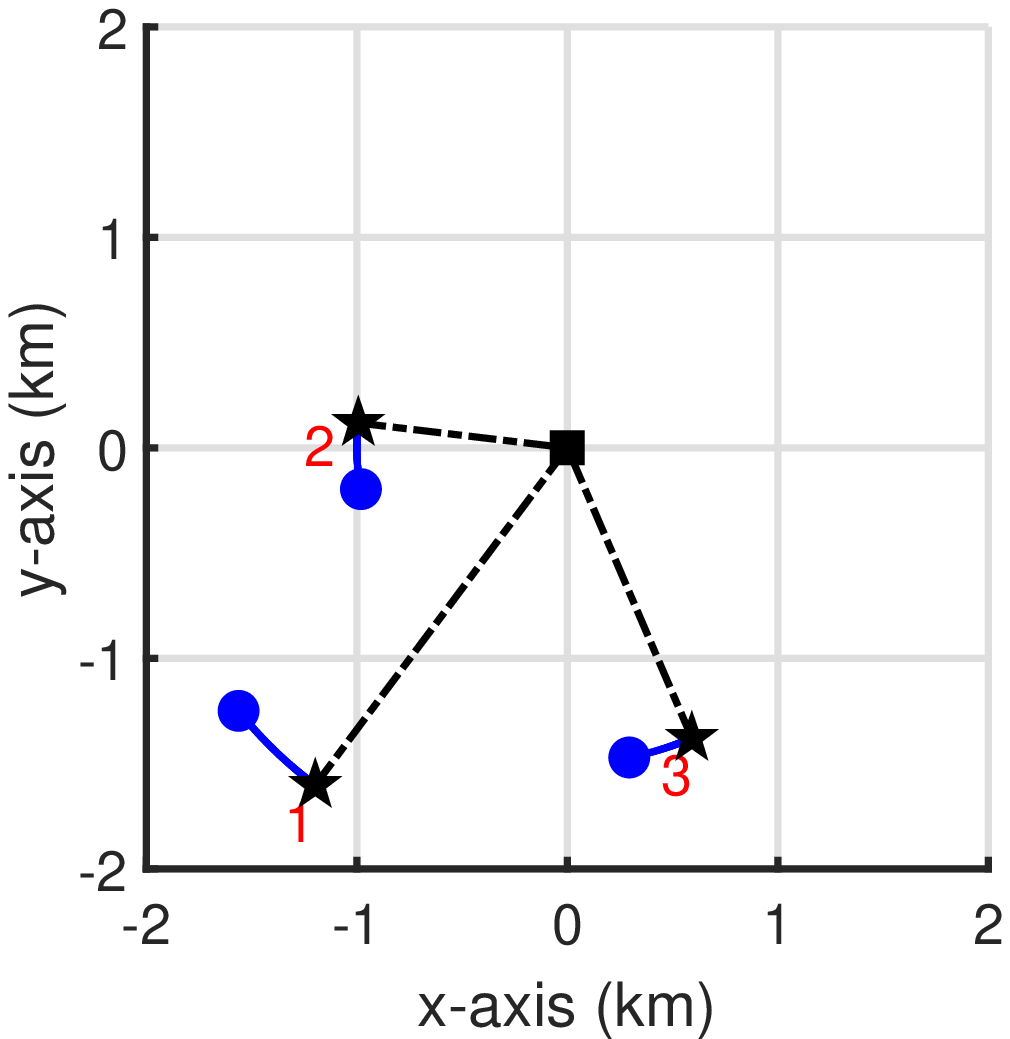}
\par\end{centering}
\caption{\label{dia5}{\small{}Optimal sensor trajectories and the final geometries
for non-uniform $d_{i}$ in example 3 and 4 }considered in subsection
{\small{}\ref{subsec:4.3.2}.}}
\end{figure}

For the analysis of optimal geometry in case of non-uniform $d_{i}$,
we have considered design problems of placing 2 and 3 sensors. In
the first case, the initial sensor positions and noise variances are
taken to be same as in example 1 but the sensor-target ranges are
taken to be $d_{1}=2$ kilometers and $d_{2}=1$ kilometers. Similarly,
in the second case, the initial sensor positions and noise variances
are same as in example 2 but the the sensor-target ranges are taken
to be $d_{1}=2$ kilometers, $d_{2}=1$ kilometers and $d_{3}=1.5$
kilometers. The optimal geometries obtained using the proposed algorithm
for these examples are given in Fig. \ref{dia5}. The minimum of the
A-optimal objective obtained are $1.99$ sq. meters and $1.33$ sq.
meters, respectively, which is equal to the minimum theoretical values
in \eqref{crlb_th}. 

\subsection{Mean Square Error (MSE) analysis}

\begin{table}[th]
\centering \caption{{\footnotesize{}COMPARISON OF THE MLE PERFORMANCE FOR DIFFERENT PLACEMENT}}
\label{t2} %
\begin{tabular}{ccc}
\hline 
{\small{}No of sensors} & {\small{}Placement} & {\small{}MSE (sq. meters)}\tabularnewline
\hline 
\hline 
 & {\small{}Random} & {\small{}$0.4371$}\tabularnewline
{\small{}$m=3$} & {\small{}Uniform} & {\small{}$0.3931$}\tabularnewline
{\small{}(uniform noise)} & {\small{}Optimal} & {\small{}$0.1384$}\tabularnewline
\hline 
 & {\small{}Random} & {\small{}$0.2763$}\tabularnewline
{\small{}$m=5$} & {\small{}Uniform} & {\small{}$0.1417$}\tabularnewline
{\small{}(uniform noise)} & {\small{}Optimal} & {\small{}$0.1082$}\tabularnewline
\hline 
 & {\small{}Random} & {\small{}$0.5830$}\tabularnewline
{\small{}$m=7$} & {\small{}Uniform} & {\small{}$0.0962$}\tabularnewline
{\small{}(uniform noise)} & {\small{}Optimal} & {\small{}$0.0867$}\tabularnewline
\hline 
 & {\small{}Random} & {\small{}$0.1767$}\tabularnewline
{\small{}$m=3$} & {\small{}Uniform} & {\small{}$0.1431$}\tabularnewline
{\small{}(non uniform noise)} & {\small{}Optimal} & {\small{}$0.1226$}\tabularnewline
\hline 
 & {\small{}Random} & {\small{}$0.0950$}\tabularnewline
{\small{}$m=5$} & {\small{}Uniform} & {\small{}$0.0876$}\tabularnewline
{\small{}(non uniform noise)} & {\small{}Optimal} & {\small{}$0.0855$}\tabularnewline
\hline 
 & {\small{}Random} & {\small{}$0.0886$}\tabularnewline
{\small{}$m=7$} & {\small{}Uniform} & {\small{}$0.0803$}\tabularnewline
{\small{}(non uniform noise)} & {\small{}Optimal} & {\small{}$0.0771$}\tabularnewline
\hline 
\end{tabular}
\end{table}

Here, we study the improvement in the target localization accuracy
using the sensor positions obtained by the proposed approach. We will
compare the MSE of the target location obtained using optimal sensor
placement given by our approach with the MSE of estimates obtained
using the uniform and random sensor placements. The target location
estimate is obtained by the maximum likelihood estimation (MLE) approach
which is implemented through a 2D grid search of the hybrid likelihood
function to arrive at the most probable target location and then performing
one step of Gauss-Newton update on the MLE estimate. The configuration
used in the simulation is given as follows: The target is considered
to be present at $[0,0]^{\text{T}}$ and the sensors are placed on
the circumference of circle with unit radius. The uniform and the
random sensor geometries are obtained by placing the sensors uniformly
and randomly over the circumference, respectively, whereas the optimal
geometry is determined using the proposed algorithm. The experiment
is conducted for noise measurements which are uncorrelated (with uniform
and non-uniform variances). The values of the MSE of the MLE, obtained
via 1000 Monte Carlo simulations, are listed in the table \ref{t2}.
From these results, we can observe that the MSE of the estimates in
case of optimal placement are smaller than the uniform and random
placements.

% \begin{table}[ht]
% \label{t_MISMATCH}
% \centering
% \caption{\\ COMPARISON OF THE MLE PERFORMANCE FOR TARGET MISMATCHING}
% \begin{tabular}{c c c c}
% \hline 
% No of sensors &Placement & MSE ($m^2$)&Bias ($m^2$)\\
% \hline
% \hline
% &Random&0.0704&0.0070\\
% m=3&Uniform & 0.0686&0.0140\\
% (correlated noise)&Optimal&0.0513&0.0134\\
% \hline
% &Random&0.0445&0.0055\\
% m=4&Uniform & 0.0440&0.0091\\
% (correlated noise)&Optimal&0.0384&0.0079\\
% \hline
% &Random&0.0067&0.0029\\
% m=5&Uniform &0.0052&0.0014 \\
% (correlated noise)&Optimal&0.0040&0.0006\\
% \hline
% &Random&0.2324&1.1414\\
% m=7&Uniform &0.0063&0.5469 \\
% (correlated noise)&Optimal&0.0021&0.3905\\
% \hline
% &Random&0.0687&0.0472\\
% m=3&Uniform & 0.0143&0.0046\\
% (uniform noise)&Optimal&0.0136&0.0012\\
% \hline
% &Random&0.0125&0.0045\\
% m=4&Uniform & 0.0119&0.0035\\
% (uniform noise)&Optimal&0.0118&0.0027\\
% \hline
% &Random&0.0119&0.0008\\
% m=5&Uniform &0.0103&0.0041 \\
% (uniform noise)&Optimal&0.0100&0.0005\\
% \hline
% &Random&0.0090&0.0040\\
% m=7&Uniform &0.0078&0.0012 \\
% (uniform noise)&Optimal&0.0077&0.0003\\
% \hline
% &Random&0.0464&0.0181\\
% m=3&Uniform & 0.0276&0.0062\\
% (non-uniform noise)&Optimal&0.0216&0.0047\\
% \hline
% &Random&0.0190&0.042\\
% m=4&Uniform & 0.0187&0.0036\\
% (non-uniform noise)&Optimal&0.0168&0.0034\\
% \hline
% &Random&0.0162&0.0018\\
% m=5&Uniform &0.0152&0.0012 \\
% (non-uniform noise)&Optimal&0.0147&0.0004\\
% \hline
% &Random&0.0080&0.0087\\
% m=7&Uniform &0.0047&0.0018 \\
% (non-uniform noise)&Optimal&0.0033&0.0013\\
% \hline
% \end{tabular}
% \end{table} 

\subsection{Performance analysis of the proposed method for mismatch in the target
location estimate}

\begin{table}[th]
\centering \caption{{\footnotesize{}COMPARISON OF THE MLE PERFORMANCE OF SENSOR PLACEMENT
FOR TARGET LOCATION MISMATCH}}
\label{tab5}%
\begin{tabular}{>{\centering}p{3cm}>{\centering}p{2.4cm}>{\centering}p{1.5cm}}
\hline 
{\small{}No of sensors} & {\small{}Target Position} & {\small{}MSE (sq. meters)}\tabularnewline
\hline 
\hline 
 & {\small{}True position} & {\small{}$0.1752$}\tabularnewline
{\small{}$m=3$} & {\small{}Shifted position 1} & {\small{}$0.1855$}\tabularnewline
{\small{}(Uniform noise)} & {\small{}Shifted Position 2} & {\small{}$0.1894$}\tabularnewline
\hline 
 & {\small{}True position} & {\small{}$0.0304$}\tabularnewline
{\small{}$m=4$} & {\small{}Shifted position 1} & {\small{}$0.0334$}\tabularnewline
{\small{}(Non-Uniform noise)} & {\small{}Shifted Position 2} & {\small{}$0.0362$}\tabularnewline
\hline 
 & {\small{}True position} & {\small{}$0.0262$}\tabularnewline
{\small{}$m=5$} & {\small{}Shifted position 1} & {\small{}$0.0305$}\tabularnewline
{\small{}(Non-Uniform noise)} & {\small{}Shifted Position 2} & {\small{}$0.0342$}\tabularnewline
\hline 
\end{tabular}
\end{table}

As discussed in the Remark \ref{rem:3}, the optimal sensor placement
is designed considering an initial estimate of the target position.
However, this initial estimate is very coarse and may not be close
to the true target position. So, in this section we study the performance
of the proposed algorithm under the mismatch in the target location
used in the design. We consider the target position used in the design
to be slightly different from the true position. The MSE of the location
estimate for the sensor placement obtained by the proposed algorithm
is compared with two such cases of target location mismatch. For this,
we consider sensors placed on the circumference of a unit circle with
target at its center. The measurements are assumed to be corrupted
by Gaussian noise. The covariance matrix of the noise is taken to
be $\mathbf{I}_{m}$ for uniform noise case and in the case of non-uniform
noise, the diagonal elements of it are randomly drawn from uniform
distribution of $\mathcal{U}(0,1)$. The target positions used in
the design are taken to be $[0.1,-0.2]^{T}$ and $[-0.3,-0.2]^{T}$
for the two mismatched cases. The MSE results for different number
of sensors are given in Table \ref{tab5}. The MSE for the target
mismatch cases are only slightly higher than that for the case which
assumes the true target position in the design, this proves the robustness
of the design under any mismatch in the design parameters.

\section{Conclusion and future work}

In this paper, a novel MM-based algorithm for optimal sensor placement
for a hybrid TOA-RSS-AOA measurement model is proposed. We first derive
the design objectives based on A, D and E-optimal criteria, and then
reformulate the design problems into saddle-point problems, which
are then solved via MM by constructing simple surrogate functions
(with closed-from optimizers) over both primal and dual variables.
The convergence, estimation efficiency and robustness of the algorithm
in designing optimal sensor locations for the hybrid model are verified
using various numerical simulations. Unlike the state-of-the-art methods,
our proposed algorithm works efficiently for both uncorrelated and
correlated noise in the measurements, and as well as for different
sensor-target ranges. Furthermore, the MSE for the optimal orientations
obtained via our proposed algorithm are found to be smaller than that
of the uniform and random placements. As a part of the future work,
we would like to model the noise in the measurements to be distance
dependent and try and develop a numerical procedure to obtain optimal
sensor locations.

\appendix{}

\subsection*{Proof of Lemma \ref{lem:1} }
\begin{IEEEproof}
Let us consider the inner minimization problem of (\ref{eqn2h}) over
the dual variable $\mathbf{\Phi}$: 
\begin{equation}
\min\limits _{\mathbf{\Phi}\succeq0}~\textrm{Tr}[\mathbf{\Phi}\textbf{H}^{T}\textbf{R}\textbf{H}]-\textbf{tr}(\sqrt{\mathbf{\Phi}}).\label{eq:3-2}
\end{equation}
The Karush-Kuhn-Tucker (KKT) condition for (\ref{eq:3-2}) is given
by: 
\begin{equation}
\textbf{H}^{T}\textbf{R}\textbf{H}-\frac{1}{2}\left(\left(\mathbf{\Phi}^{*}\right)^{-1/2}\right)=\mathbf{0}.\label{eq:4-1}
\end{equation}
Solving (\ref{eq:4-1}), we get the following optimal value:
\begin{equation}
\mathbf{\Phi}^{*}=\frac{1}{4}\left(\textbf{H}^{T}\textbf{R}\textbf{H}\right)^{-2}.
\end{equation}
Substituting back $\mathbf{\Phi}^{*}$ in (\ref{eqn2h}), we get:
\begin{equation}
\begin{array}{ll}
\max\limits _{\textbf{H},\textbf{J}\in\mathcal{D}}~-\textrm{Tr}(\frac{1}{4}\textbf{H}^{T}\textbf{R}\textbf{H})^{-1}, & \;\textrm{s.t. }\textbf{H}=\begin{bmatrix}\textbf{J}\\
\textbf{J}\textbf{U}
\end{bmatrix}.\end{array}\label{eq:7-1}
\end{equation}
Leaving out the constant $\frac{1}{4}$ and converting (\ref{eq:7-1})
into a minimization problem gives the original problem (\ref{eqn1h}).
Hence, the equivalence of (\ref{eqn1h}) and (\ref{eqn2h}) is proved.
\end{IEEEproof}

\end{document}